\documentclass[a4paper,reqno]{amsart}
\usepackage[text={5.0in,8in},centering]{geometry}

\usepackage[usenames,dvipsnames,svgnames,table]{xcolor}
\definecolor{citegreen}{rgb}{0.00,0.70,0.30}
\usepackage[colorlinks=true,
            linkcolor=blue,
            urlcolor=blue,
            citecolor=blue]{hyperref}

\usepackage{hyperref}

\usepackage{amsrefs}

\usepackage{amssymb}
\usepackage{mathrsfs}
\usepackage{mathtools}

\usepackage{xspace}

\usepackage{stmaryrd}

\usepackage[title]{appendix}

\usepackage{pgf}
\usepackage{tikz}
\DeclareMathAlphabet{\mathpzc}{OT1}{pzc}{m}{it}

\usepackage{epsfig}
\usepackage{enumerate}
\usepackage{graphicx}

\numberwithin{equation}{section}
\theoremstyle{plain}
\newtheorem{theorem}{Theorem}[section]
\newtheorem{prop}[theorem]{Proposition}
\newtheorem{lemma}[theorem]{Lemma}

\newtheorem{cor}[theorem]{Corollary}
\theoremstyle{remark}
\newtheorem{remark}{Remark}[section]

\newtheorem*{quest*}{Question}
\newtheorem*{remark*}{Remark}
\theoremstyle{remark}

\theoremstyle{definition}
\newtheorem{definition}{Definition}[section]
\newtheorem*{definition*}{Definition}
\newtheorem*{notation*}{Notation}
\newtheorem*{notations*}{Notations}
\providecommand{\B}{\mathbf}

\providecommand{\D}{\mathbb}

\providecommand{\R}{\mathrm}

\newcommand{\eu}{\mathrm{e}}
\newcommand{\ii}{\mathrm{i}}

\def\ball{\mathrm{B}}
\def\bball{\mathbf{B}}
\def\bballn{\mathbf{B}^{(\fn)}}
\def\bballN{\mathbf{B}^{(\fN)}}
\def\bballNj{\mathbf{B}^{(\fN)}_{j}}

\def\mytimes{\operatornamewithlimits{\hbox{\huge$\times$}}}

\DeclareMathOperator{\dist}{dist}

\DeclareMathOperator{\card}{card}

\DeclareMathOperator*{\essup}{ess\,sup}
\DeclareMathOperator*{\supp}{supp}

\DeclareMathOperator{\inneq}{{}_{\neq}\!{\in}}

\DeclareMathOperator{\one}{\mathbf{1}}

\DeclareMathOperator{\diam}{{\rm diam}}
\DeclareMathOperator{\Unif}{{\rm Unif}}

\def\Sep#1{{{\rm\mathbf{Sep}}\left[#1\right]}}
\def\Sepb#1{{{\rm\mathbf{Sep}}\big[#1\big]}}

\def\Nj{{N_j}}

\def\fh{\mathfrak{h}}

\def\fs{\mathfrak{s}}

\def\fBa{\mathbf{\mathfrak{a}}}
\def\fBA{\mathbf{\mathfrak{A}}}

\def\taujl{\tau_{j,l}}

\def\rPjl{{\R{P}_{j,l}}}

\def\bco{\begin{cor}}
\def\eco{\end{cor}}

\def\lr#1{{\langle #1\rangle}}

\def\SigmaNj{\Sigma^{\tN_j}}

\def\Sgomth#1#2{{\Sigma(\BH_{\bball_{#1}(#2)}(\om,\th))}}

\def\icirc{{i_\circ}}

\def\bcEN{\boldsymbol{\mathcal{E}}^{N}}

\def\vBx{\vec{\B{x}}}

\def\cdott{\cdot\,}

\def\cond{\,\big|\,}

\def\rc{\mathrm{c}}

\def\lam{{\lambda}}

\def\tomega{\tilde{\omega}}
\def\tOmega{\tilde{\Omega}}
\def\th{\vartheta}
\def\Th{{\rm \Theta}}
\def\eps{\epsilon}

\def\ffi{\varphi}

\def\prth#1{{\DP^{\Th}\left\{\,#1\,\right\}}}
\def\prthp{{\DP^{\Th}}}

\def\prTh#1{{\DP^{\Th}\left\{\,#1\,\right\}}}

\def\prtild{{\widetilde{\DP}}}

\def\prtil#1{{\widetilde{\DP}\left\{\,#1\,\right\}}}

\def\esmth#1{\D{E}^{\Theta}\left[\, #1\, \right]}

\def\hth{{{\widehat{\th}}}}

\def\thnk{{\th_{n,k}}}

\def\ffink{{\ffi_{n,k}}}

\def\Cnk{{C_{n,k}}}

\def\Thinf{{\Th^{(\infty)}}}

\def\Thzero{{\Th^{(0)}}}

\def\tA{{ \widetilde{A} }}

\def\hk{{\widehat{k}}}

\def\tNL{{\widetilde{N}(L)}}
\def\tN{{\widetilde{N}}}
\def\tn{{\widetilde{n}}}

\def\DIV{{\rm\textbf{(DIV)}}\xspace}
\def\LVB{{\rm\textbf{(LVB)}}\xspace}
\def\LVBU{{\rm\textbf{(LVBU)}}\xspace}

\def\UPA{{\rm\textbf{(UPA)}}\xspace}
\def\RCM{{\rm\textbf{(RCM)}}\xspace}

\def\ER{$E$-R\xspace}

\def\EmS{$(E,m)$-S\xspace}
\def\EmNS{$(E,m)$-NS\xspace}

\def\ER{$E$-R\xspace}
\def\ENR{$E$-NR\xspace}

\def\rhoL{\rho^{(L)}}
\def\brho{{\boldsymbol{\rho}}}

\def\hBx{\widehat{\Bx}}
\def\hBy{\widehat{\By}}

\providecommand{\bS}[1]{\boldsymbol{#1}}

\def\Brho{\boldsymbol{\rho}}

\def\BA{\mathbf{A}}

\def\BG{\mathbf{G}}
\def\BH{\mathbf{H}}

\def\BK{\mathbf{K}}

\def\BT{\mathbf{T}}
\def\BU{\mathbf{U}}

\def\BV{\mathbf{V}}
\def\BW{\mathbf{W}}

\def\Bc{\mathbf{c}}
\def\Bn{\mathbf{n}}
\def\Bx{\mathbf{x}}
\def\By{\mathbf{y}}
\def\Bu{\mathbf{u}}

\def\Bz{\mathbf{z}}
\def\BX{\mathbf{X}}
\def\BY{\mathbf{Y}}
\def\BZ{\mathbf{Z}}

\def\Bzero{\mathbf{0}}

\def\BGamma{{\boldsymbol{\Gamma}}}
\def\BDelta{{\boldsymbol{\Delta}}}

\def\Bpsi{{\boldsymbol{\psi}}}
\def\Bphi{{\boldsymbol{\phi}}}
\def\Bchi{{\boldsymbol{\chi}}}
\def\BomN{{\boldsymbol{\omega}^{(\mathfrak{N})}}}

\def\BLam{\mathbf{\Lambda}}

\def\BPsi{{\bS{\Psi}}}
\def\BPhi{{\bS{\Phi}}}

\def\Const{{\rm{Const}}}

\def\tn{{\widetilde{n}}}

\def\DC{\D{C}}

\def\DP{\D{P}}
\def\DR{\D{R}}
\def\DZ{\D{Z}}
\def\DN{\D{N}}
\def\DT{\D{T}}
\def\DQ{\D{Q}}

\def\cB{\mathcal{B}}

\def\csB{\mathscr{B}}

\def\csV{\mathscr{V}}

\def\cC{\mathcal{C}}

\def\cE{\mathcal{E}}

\def\cG{\mathcal{G}}
\def\cH{\mathcal{H}}

\def\cJ{\mathcal{J}}

\def\fk{\mathfrak{k}}
\def\fN{\mathfrak{N}}
\def\fn{\mathfrak{n}}

\def\cK{\mathcal{K}}
\def\cL{\mathcal{L}}
\def\cM{\mathcal{M}}

\def\csP{\mathscr{P}}

\def\cT{\mathcal{T}}
\def\cU{\mathcal{U}}

\def\cZ{\mathcal{Z}}

\def\cV{\mathcal{V}}

\def\hlam{\hat{\lambda}}

\def\rA{{\R{A}}}

\def\bballj{\mathrm{\mathbf{B}}_{L_j}}
\def\bballjone{\mathrm{\mathbf{B}}_{L_{j+1}}}

\def\mbad{$m${\rm-bad}\xspace}
\def\mgood{$m${\rm-good}\xspace}

\def\sparsez{{\rm \textbf{Sparse($\mathbf{0}$)}}\xspace}
\def\sparseL#1{{\rm \textbf{Sparse($\mathbf{#1}$)}}\xspace}
\def\sparsepar#1{{\rm \textbf{Sparse($\mathbf{L_{#1}}$)}}\xspace}
\def\sparsej{{\rm \textbf{Sparse($\mathbf{L_{j}}$)}}\xspace}
\def\sparsejone{{\rm \textbf{Sparse($\mathbf{L_{j+1}}$)}}\xspace}

\def\hX{\hat{X}}

\def\rd{{\R{d}}}
\def\rP{{\R{P}}}

\def\hBx{\hat{\mathbf{x}}}

\def\be{\begin{equation}}
\def\ee{\end{equation}}
\def\ba{\begin{array}{l}}
\def\ea{\end{array}}
\def\bal{\begin{aligned}}
\def\eal{\end{aligned}}

\def\ble{\begin{lemma}}
\def\ele{\end{lemma}}

\def\bpr{\begin{prop}}
\def\epr{\end{prop}}

\def\bre{\begin{remark}}
\def\ere{\end{remark}}

\def\btm{\begin{theorem}}
\def\etm{\end{theorem}}

\def\bde{\begin{definition}}
\def\ede{\end{definition}}

\def\ffi{{\varphi}}

\def\fB{\mathfrak{B}}
\def\fF{\mathfrak{F}}
\def\fG{\mathfrak{G}}
\def\tfF{\widetilde{\mathfrak{F}}}
\def\fS{\mathfrak{S}}
\def\fs{\mathfrak{s}}

\def\om{{\omega}}
\def\Om{{\Omega}}

\def\eps{\epsilon}

\def\Lam{{\Lambda}}
\def\lam{{\lambda}}

\def\bcZ{{\boldsymbol{\mathcal{Z}}}}
\def\bcZN{{\boldsymbol{\mathcal{Z}^{\fN}}}}
\def\bcZNge{{\boldsymbol{\mathcal{Z}_{\ge}^{\fN}}}}

\def\bcEN{{\boldsymbol{\mathcal{E}^{(\fN)}}}}

\def\pr#1{\D{P}\left\{\,#1\,\right\}}
\def\esm#1{\D{E}\left[\, #1\, \right]}
\def\pt{\partial}
\def\half{\frac{1}{2}}

\def\quart{\frac{1}{4}}

\def\nb#1{{ \langle #1 \rangle}}

\def\truc#1#2#3{\smash{\mathop{\,\, #1 \,\, }\limits^{#2}_{#3}}}

\def\tto#1{\smash{\mathop{\,\,\,\, \longrightarrow \,\,\,\, }\limits_{#1}}}

\def\myset#1{{\left\{\,#1\,\right\}}}

\def\bra#1{{ \langle {#1}| }}
\def\ket#1{{ | {#1} \rangle }}

\def\mymax#1{{ \truc{\max} {} {#1}}}

\def\diy{\displaystyle}

\def\Uone{{\rm\textbf{(INT)}}\xspace}

\definecolor{redd}{rgb}{0.95,0.2,0.2}
\definecolor{gris}{rgb}{0.9,0.9,0.9}

\definecolor{cmgray}{rgb}{0.7,0.7,0.7}

\definecolor{cmblue}{rgb}{0.2,0.5,0.8}

\begin{document}

\title[Uniform $N$-particle Anderson localization]
{Uniform $N$-particle Anderson localization
\\and unimodal eigenstates
\\in deterministic disordered media
\\without induction on the number of particles}

\author[V. Chulaevsky]{Victor Chulaevsky}


\address{D\'{e}partement de Math\'{e}matiques\\
Universit\'{e} de Reims, Moulin de la Housse, B.P. 1039\\
51687 Reims Cedex 2, France\\
E-mail: victor.tchoulaevski@univ-reims.fr}

\date{}
\begin{abstract}
We present the first rigorous result on Anderson localization for
interacting systems of quantum particles subject to a deterministic (e.g.,
almost periodic) disordered external potential. For a particular class of
deterministic, fermionic, Anderson-type  Hamiltonians on the lattice of an arbitrary dimension,
and for a large class of underlying dynamical systems generating the external potential,
we prove that the spectrum is pure point, all eigenstates are unimodal
and feature a uniform exponential decay.
In contrast to all prior mathematical works on multi-particle Anderson localization,
we do not use the induction on the number of particles.
\end{abstract}

\maketitle

\section{Introduction. The model and the main results.}\label{intro}

We study spectral properties of finite-difference operators, usually called
discrete Schr\"{o}dinger operators (DSO), arising as Hamiltonians of
$\fN$-particle\footnote{We denote the number of particles by the Gothic letters $\fn, \fN$,
since the letters $n, N$ are used in our paper for other purposes.}
fermionic quantum systems on $\DZ^d$ with a nontrivial interaction of infinite range,
subject to the common external potential. Specifically, we consider the Hamiltonians of the form
\be\label{eq:Np.DSO}
\bal
\BH(\om;\th) &= \BH_0 + \BV(\om;\th) + \BU(\Bx)
\\
& =  \sum_{j=1}^\fN \Big[ H_{0;j} + gV(x_j;\om;\th) \Big] + \sum_{1\le j < k \le \fN} U(|x_j - x_k|)
\eal
\ee
restricted to the subspace $\cH_-^{\fN}$ of antisymmetric functions in the Hilbert space
$\cH^\fN = \ell^2((\DZ^d)^\fN)$
of square-summable complex functions $\Bpsi: (\DZ^d)^\fN \to \DC$. A self-consistent representation
of $\BH(\om;\th)$, not referring to the configuration space $(\DZ^d)^\fN$ of distinguishable particles,
is given in Eqn. \eqref{eq:def.fermionic.H}.

In some formulae will appear the notations like $\BH^{(N)}$; we stress that the superscript $N$
does not have the meaning of the number of particles, but refers to an auxiliary
approximation procedure. As explained below, we do not use the induction of the number of particles,
so the integer $\fN$ is assumed to be chosen and then fixed; its value can be arbitrary.

The structure of the external potential $x \mapsto gV(x;\om;\th)$, acting on each particle, is
described in Sect.~\ref{sec:param.ensembles} (in particular, cf. subsection \ref{ssec:randelettes}).

For the first time in mathematical literature, we establish Anderson localization for a system
of interacting quantum particles in a \emph{deterministic} (e.g., quasi-periodic) external potential.
In all  prior rigorous works on multi-particle Anderson localization, it
was assumed that the external random potential has independent values or features a fast decay
of correlations at infinity (Rosenblatt strong mixing condition).
Comparing the present paper with earlier
results on multi-particle localization, we would like to stress
the following.

\vskip1mm

$\blacklozenge$ All earlier mathematical works on $\fN$-particle Anderson localization used
the induction on the number of particles, which was introduced in \cite{CS09a} in the particular case of $2$-particle
systems, as the transition from $\fN=1$ to $\fN=2$, and later generalized in \cite{CS09b}
(in the framework of the Multi-Particle Multi-Scale Analysis, MPMSA) and in \cite{AW09a}
(with the help of the Multi-Particle variant of the Fractional Moment Method, MPFMM).
In the present paper, we develop a different approach where the induction in $\fN$
becomes unnecessary. This results in a streamlined proof of multi-particle localization
(more precisely, the analytical component thereof). However,
just as in our work \cite{C13b} where single-particle deterministic Anderson models are considered,
one has to establish first rather strong, uniform lower bounds on the spectral spacings for
the Hamiltonians at hand.
Such bounds can be considered as  deterministic analogs of the celebrated Wegner estimate
\cite{W81} (cf. \cite{CS08,CBSS09} for its multi-particle counterparts).

$\blacklozenge$ Another distinction of the present paper from earlier works \cite{CS09b,AW09a,CBS11}
on random multi-particle  Anderson Hamiltonians,
and more
recent papers by Klein and Nguyen \cite{KN13a,KN13b}, is that we prove uniform decay bounds
on the decay of eigenfunctions and of eigenfunction correlators in the genuine norm-distance
on the lattice, while the above mentioned works operated, explicitly or implicitly, with
the so-called Hausdorff distance. As a result, efficient localization bounds could be proven
only in the actually infinite lattice, but not in an [arbitrarily large] finite volume.
Yet, in the applications to physical models, localization is to be studied in a finite
region of the physical space.

$\blacklozenge$ We allow for the interaction potentials decaying only slightly faster than polynomially
(cf. \eqref{eq:decay.U.intro}). This is the first rigorous result on multi-particle Anderson localization
in presence of an interaction decaying much slower than at exponential or sub-exponential rate
$r\mapsto \eu^{-r^\zeta}$, $\zeta>0$.

\vskip3mm

Building on the techniques
developed in \cite{C11c,C13b} for $1$-particle systems, we thus complement
the existing rigorous methods of the multi-particle Anderson localization theory
(MPMSA and MPFMM) with a new approach, providing for a particular class of models with nontrivial interaction
much stronger results than for random Hamiltonians: uniform exponential localization
and unimodality of all eigenstates,
observed in the conventional, single-particle theory only in systems subject to an almost-periodic
external potential (cf. \cite{FGP84}, \cite{BLS83}, and more recently \cite{DG10,DG12}).

There is a considerable overlap of the present text with Ref. \cite{C13b}, mainly in the
analysis of spectral spacings required for the proof of uniform localization.
Unfortunately, we cannot simply refer to the technical results proved in \cite{C13b},
for they have to be adapted to the multi-particle setting.
In addition, the multi-particle model with a nontrivial interaction
brings in its own lot of technical problems. This explains why the present
manuscript is substantially longer than \cite{C13b}.

It is essential for our method to have the $\fN$-particle Hamiltonian $\BH(\om;\th)$ restricted
to a proper subspace of the symmetric group $\fS_\fN$ acting in $\cH^\fN$ by permutations of the
particle positions.
In the entire space $\cH^\fN$, there are unavoidable "resonances"
which render impossible the genuine \emph{uniform} localization of \emph{unimodal}  eigenstates.
On the other hand, choosing the fermionic subspace (rather than the bosonic one) is merely a matter of convenience.

In fact, one can  prove Semi-Uniform (exponential) Localization of Eigenfunctions
(usually referred to as SULE),
directly in the entire Hilbert space of \emph{distinguishable} quantum particles,
for a richer class of deterministic potentials than the one considered here.
This requires an adaptation of the method developed in our earlier works with
Yuri Suhov \cite{CS09a,CS09b} (induction on the number of particles), which is technically
more involved. We plan to make such an adaptation in a separate work.

In Eqn. \eqref{eq:Np.DSO},
$H_{0;j}$ is the kinetic energy operator acting on the $j$-th particle, $x_j\in\DZ^d$
is the position of the $j$-th particle, and $U(r)$, $r\ge 0$, is the two-body interaction potential,
which we assume decaying as follows
\be\label{eq:decay.U.intro}
\forall\, r \ge 1\quad |U(r)| \le  r^{- 2B \ln r},
\ee
where $B = 400 b A^2/\ln 2$  (cf. \eqref{eq:def.B}) with the same value $A>0$ as in \UPA
and $b>0$ as in \eqref{eq:def.an.b}.

One could
assume $\BU$ to be a fairly general, symmetric $k$-body interaction, $k\ge 2$; this would not require
any substantial modification of the proofs, while making notation more cumbersome.

For clarity, we always assume the single-particle kinetic energy operators $H_{0;j}$ to be replicas
of the second-order discrete Schr\"{o}dinger operator (DSO) of the form
\be\label{eq:DSO.intro}
(H_0 f)(x) = \sum_{y:\, \|y-x\|=1} f(y) .
\ee
It is not difficult, however, to consider a more general, finite-difference operator
$H_0$.

In contrast to the usual approach of the Anderson localization theory for Hamiltonians with
random (e.g., IID) random potential,
we consider a \emph{parametric family} of external potentials, generated by a "hull" function
\be
v: \Om\times \Th \to \DR
\ee
on the Cartesian product of the phase space $(\Om,\fF,\DP)$ of a dynamical system
\be
T: \, \DZ^d \times\Om \to\Om
\ee
with discrete time $\DZ^d$,
and of an auxiliary parameter space which we endow with the structure of a probability
space $(\Th,\fB,\prthp)$. Specifically, we set
\be
V(x;\om;\th) = v(T^x\om; \th),
\ee
so that the dynamical system, acting on $\om\in\Om$, leaves invariant the parameter
$\th\in\Th$. The structure of the "hull" function $v$ is discussed in detail in
Sections \ref{ssec:randelettes} and \ref{sec:randelettes.and.sep.bounds}.
Summarising, one can say that
we expand the hull $\om\mapsto v(\om)$ in a convergent series, $v(\om)=\sum_{n\ge 0} c_i \phi_i(\om)$,
and consider the expansion coefficients $c_i$ as independent parameters.
Introducing the parameter space makes possible the exclusion of
"unwanted" hulls essentially in the same way as
one excludes a subset of random samples $\om\in\Om$ in the localization theory of random operators.
It is this specific construction which allows us to
prove the main result on genuinely uniform Anderson localization for typical values
of the expansion coefficients (see Theorem \ref{thm:Main} in Section \ref{sec:Main.res}).
We encapsulate the main requirement for the underlying dynamical system
generating the deterministic random potential in one mild condition -- "Uniform
Power-law Aperiodicity" (\UPA; cf. \eqref{eq:cond.UPA} in Section \ref{sec:UPA}).

Putting the factor $g$ in front of the external potential energy
in \eqref{eq:Np.DSO}
is a natural way to control the amplitude of the (deterministic) disorder.
Since  $V$, as well as $U)$ can be of arbitrary sign or sign-indefinite,
we always assume that $g>0$, for notational brevity.

A rich and  interesting class of \emph{quasi-periodic} potentials, e.g., in one dimension,
is obtained when $\Om$ is the torus $\DT^1$
endowed with the  Haar measure $\DP$, and  the  dynamical system on $\Om$ is  given by
$
T^x:\, \om  \mapsto  \om + x \alpha, \;\; \om\in \DT^1,
$
and $\alpha\in\DR\setminus\DQ$.
As is well-known, this dynamical system is ergodic. Taking a function $v:\DT^1 \to \DR$, we can define an ergodic family of quasi-periodic potentials $V:\DZ\to \DR$ by $V(x;\om) := v(T^x \om)$. Multi-dimensional quasi-periodic potentials on $\DZ^d$ can be constructed in a similar way (with the help of $d$ incommensurate frequency vectors $\alpha^{j}\in\DR^\nu, j=1, \ldots, d$).
In the  case where  $v(\om) = g\cos (2\pi \om)$, $g\in\DR$,  $\alpha\in\DR\setminus \DQ$,
the DSO $H(\om)$ with the potential $V(x;\om) = v(T^x\om)$
is called  Almost Mathieu or Harper's operator.

\vskip1mm
\noindent
\textbf{A terminological remark: } Following the tradition of the harmonic analysis on compact
abelian groups, one often assumes some regularity of
a function $v:\Om\to\DR$ on the group $\Om$, when qualifying it as "almost-periodic". We employ
a more liberal terminology, focusing on the properties of the dynamical system at hand,
$T:\DZ^d\times\Om\to\Om$, and making abstraction, e.g., of the continuity of the hulls $v$.

\vskip1mm

Sinai \cite{Sin87}  and Fr\"{o}hlich et al. \cite{FSW90}
proved Anderson localization
for a class of the (single-particle)
DSO with the cosine-like potential; more precisely, the hull $v:\DT^1\to\DR$
was assumed to be of the class $\cC^2(\DT^1)$ and have exactly two extrema, both non-degenerate.
Operators with several basic frequencies (i.e., $\om\in\DT^\nu$, $\nu >1$) were studied
in \cite{CSin89} ($\nu=2$),
and later in a cycle of works by Bourgain, Goldstein and Schlag, for various dynamical
systems on a torus $\Om=\DT^\nu$, $\nu\le 2$,
where the "hull" $v(\om)$ was assumed analytic; see, e.g., \cite{BG00,BS00,BGS01}.
Chan \cite{Chan07} used a parameter exclusion technique (different from ours) to establish the
localization for quasi-periodic operators with sufficiently non-degenerate hull $v\in \cC^3(\DT^1)$.

\vskip1mm

Our model features unusually strong localization properties, similar to those of the celebrated
Maryland model, discovered and studied by the team of physicists Fishman et al. \cite{FGP84}.
The potential in the Maryland model is  quasi-periodic and generated by the analytic hull
$$
\om \mapsto g \, \tan \pi\om, \;\; \om\in\DT^1 \cong [0,1) \subset \DR \hookrightarrow \DC,
$$
which admits a meromorphic continuation to the complex plane. Its restriction to $\DR$
is strictly monotone on the period (between two consecutive poles),
and this ultimately results in complete absence of ``resonances''
between distant sites in the lattice $\DZ^d$. In turn, this gives rise
to the exponentially localized eigenstates which are unimodal, i.e., cannot have multiple "peaks".

The notion of a "peak" becomes meaningful
for the disorder amplitude $|g|\gg 1$: in this case, the Maryland operator has an
orthonormal eigenbasis
of exponentially fast decaying eigenfunctions $\psi_x$, labeled in a non-ambiguous and natural way by
the points $x\in\DZ^d$ so that
$$
\begin{aligned}
\inf_{x\in \DZ^d} \; |\psi_x(x)|^2 &\ge 1 - C|g|^{-1/2} > \half .
\end{aligned}
$$
In other words, for $|g|\gg 1$,
the eigenbasis for $H(\om)$ is a small-norm perturbation of the
standard delta-basis in $\ell^2(\DZ^d)$; this would be, of course,  an event of probability 0 for random Anderson Hamiltonians.

In the Almost Mathieu model and, more generally, in the class of models studied by Sinai
\cite{Sin87}  and Fr\"{o}hlich et al. \cite{FSW90}, with a single-frequency, cosine-like potential,
typical eigenfunctions have multiple peaks; the set of locations of peaks of an eigenfunction $\psi$
was called in \cite{Sin87} \emph{the essential support} of $\psi$. The multiple peaks are, in fact,
inevitable in the deterministic (e.g., quasi-periodic) models with the potential generated by
a continuous hull $v$ on the phase space of the underlying dynamical system; they have
a topological nature. In the random Anderson models with, say, IID potentials, a probabilistic mechanism
is responsible for the occurrence of multiple peaks, with probability one. These observations clearly
set apart the deterministic models with "rigid" (in particular, quasi-periodic) potentials where
exceptional mechanisms prevent the eigenfunctions from having multiple peaks. In the author's opinion, these mechanisms, observed in  single-particle (cf. \cite{C13b}) and in multi-particle
deterministic Anderson models (considered here)
are less robust than those which give rise to the \emph{Semi-Uniform}
Localization of Eigenfunctions  (cf. \cite{C12a}).

Damanik and Gan proved uniform localization for a class of (single-particle)
Schr\"{o}dinger operators with the so-called \emph{limit-periodic} potentials
in $\DZ^1$ \cite{DG10}, and more recently in $\DZ^d$, $d\ge 1$ \cite{DG12}.
Recall that a function $f:\DZ^d \to \DR$ (resp., on $\DR^d$) is called limit-periodic
if it is the uniform limit of a sequence of periodic functions $f_n:\DZ^d\to\DR$
(resp., on $\DR^d$). Such an exceptionally strong form of convergence of the approximants $f_n$ suggests
that spectral properties of the limiting operator would resemble that of the periodic
approximants; indeed, this was proved for several classes of limit-periodic Schr\"{o}dinger
operators (mainly on $\DR^1$), under the condition of sufficiently fast, uniform convergence
$\|f_n - f\|_\infty \to 0$.
It is to be stressed that the rate of convergence is to be related to the rate of growth
of the periods of the approximants $f_n$ (cf., e.g., \cite{C79,Mos81,AS82,PT84}).
On the other hand, it was shown in our earlier work \cite{MC84}, in the framework of one-dimensional
Schr\"{o}dinger operators with limit-periodic potentials,
$
H = -\frac{d^2}{dx^2} + V(x),
$
with
$$
V(x) = \sum_{n\ge 1} a_n v_n(T_nx), \;\; v_n(x+T_n) \equiv v_n(x),
$$
that a sufficiently rapid growth of the periods $T_n$ gives rise to rapidly growing
(or rapidly decaying)  solutions of $H \psi = E \psi$, for "generic" periodic components $v_n$.
The uniform localization proven in \cite{DG10,DG12} also requires the periods $T_n$ (which are vectors when
$d\ge 2$) to grow fast enough, once the convergence rate $\|f_n - f\|_\infty\to 0$ has been fixed.

\vskip1mm

Another particularity of the Maryland model, rigorously proven in independent
mathematical works by Pastur--Figotin \cite{FiP84} and Simon \cite{Sim85},
is the \emph{non-perturbative} complete exponential localization: it occurs for
any, arbitrarily small amplitude of disorder $|g|>0$. With the exception for this particular feature,
the "unimodal", or "uniform exponential" localization was extended by Bellissard, Lima and Scoppola
\cite{BLS83}
to the class of meromorphic hulls with a real period, strictly monotone on the period.

The class of deterministic Anderson
models considered in this paper features the same complete unimodality
of the eigenbasis, i.e.,  genuinely uniform decay of all eigenfunctions, and not just semi-uniform
(often referenced to as SULE property: Semi-Uniformly Localized Eigenfunctions). It is to be emphasized
that our class of models also has significant differences from the Maryland and the BLS-type models:

\begin{enumerate}
  \item The class of the underlying dynamical systems, representing the disorder from the traditional point of view, is not limited to quasi-periodic or, more generally, almost-periodic systems. This is explained by the fact that the "dynamical disorder" plays a subordinate, indeed minor role
      in the physics of Anderson localization phenomenon, while the dominant role is given to the "parametric disorder", responsible for the decay of the eigenfunctions.

  \item The uniform decay of eigenfunctions occurs for \emph{all} phase points of
      the dynamical system, and not just for Lebesgue-almost all, as in many quasi-periodic systems.
      On the other hand, it occurs only for a subset of the parameter
      set, labeling the hulls. The measure of the excluded subset decays as $|g|\to\infty$
      but remains positive (at least, as far as the rigorous proofs are concerned) for any finite $g$.
      In other words, we prove localization for a.e. parameter $\th\in\Th$
      and for all $\om\in\Om$, but with $g \ge g_0(\th)$.

\end{enumerate}

\vskip1mm
$\blacklozenge$ In this work, as in \cite{C11c},
we often use the term \emph{random}, sometimes putting it in quotes, and this might create the
\textbf{illusion} that the operators with deterministic -- e.g., quasi-periodic -- potentials,
considered here,
are somehow perturbed by a masterly hidden random noise.
We stress that the external potential always remains deterministic, with stochastic
properties induced exclusively by the underlying dynamical system.
For example, if $\{T^x, x\in\DZ^d\}$ is generated by incommensurate shifts of the torus,
the obtained potentials are \textbf{genuinely quasi-periodic}.
It is true, however, that many \emph{techniques} used in the proof of localization come from the
conventional theory of random Anderson Hamiltonians, and \emph{this is} one of the main points
of our approach, where the probabilistic vocabulary, used in the context of the parametric
disorder (or, rather, parametric freedom), proves instrumental.
\vskip1mm

\vskip1mm

\section{Indistinguishable particles and fermionic Hamiltonians }
\label{sec:sym.power.graph}

\subsection{Some notational conventions}

We denote by $[[a,b]]$ the integer intervals $\{a, a+1, \ldots, b\}$, with $a,b\in\DZ$, $a \le b$.

For $t\in\DR$, $\lfloor t \rfloor$  stands for integer part of $t$, i.e., the largest integer $n \le t$.

As usual, we set $s \vee t := \max\{s, t\}$.

The symmetric difference
of arbitrary sets $A, B$, i.e., $(A\cup B) \setminus (A \cap B)$, is denoted by $A\ominus B$, since
the symbol $\Delta$ is reserved for the graph Laplacians.

As a rule, we use boldface letters to denote "multi-particle" objects. As was already said,
the number of particles will be usually denoted by Gothic letters $\fN, \fn$.

Given a finite interval $I \subset \DR$, we denote by $\Unif(I)$ the uniform probability
distribution in $I$.

To avoid cumbersome formulae, we will sometimes use notation $a,b\inneq A$,
meaning that $a$ and $b$ are two distinct elements of the set $A$.

\subsection{Symmetric powers of graphs}
\label{ssec:sym.powers}

In contrast with earlier works \cite{CS09a,CS09b}, \cite{AW09a}, where the
quantum particles were considered as distinguishable (i.e., the spectral analysis of
$\fN$-particle operators was carried out in the entire tensor power $\cH_1^{\otimes \fN}$ of
the 1-particle quantum state space $\cH_1 = \ell^2(\DZ^d)$), we adopt here a more traditional point of view
of quantum mechanics and consider particles indistinguishable. Furthermore,
quantum states $\Psi(x_1, \ldots, x_\fN)$ of an $\fN$-particle system must be
either symmetric in $x_1, \ldots, x_\fN$ (Bose--Einstein quantum statistics) or
antisymmetric (Fermi--Dirac quantum statistics). While the localization phenomena (pure point spectrum,
exponential decay of eigenfunctions and the strong dynamical localization) can be established in the
entire Hilbert space $\cH_1^{\otimes \fN}$, restricting the spectral problem to any of the eigenspaces of
(the unitary representation of) the symmetry group $\fS_\fN$ acting
by permutations of the coordinates allows one to simplify the scaling analysis. Indeed, the analysis of
the so-called
resonances\footnote{In the framework of the multi-particle MSA, we define this notion
in Sect.~\ref{ssec:res.sparse}.}
is the crucial component of the MSA (or the FMM), but in a system with a spatial symmetry (e.g., related
to the particle permutation group $\fS_\fN$) a "resonance" occurring near some
locus $(x_1, \ldots, x_\fN)$ automatically occurs near all the loci of the orbit of
$(x_1, \ldots, x_\fN)$ by the symmetry group. These multiple "fantom resonances" are mere artifacts
of the language of distinguishable particles, yet a straightforward application of the MSA
technique is bound to take  into account all these "resonances". Such a difficulty can be avoided,
when the spectral analysis is performed only in an eigen-subspace of the symmetry group.

We choose the fermionic systems; this is not crucial to our technique but results in some minor
simplifications.

\vskip1mm
Recall the conventional construction of a symmetric power of a (locally finite) graph.
Given a graph $(\cZ, \cE)$ with the vertex set $\cZ$ and the edge set $\cE$,
and an integer $\fN\ge 2$, consider the subset of $\fN$-tuples
of pairwise distinct points in the $\fN$-th cartesian power of $\cZ$,
$$
\bcZN = \big\{ \{x_1, \ldots, x_\fN\}:  x_j\in \cZ^\fN, j\in[1,\fN],\; \card\{x_1, \ldots, x_\fN\} = \fN \big\}.
$$
To remind, or stress, that the particle positions in a configuration $\Bx=\{x_1, \ldots, x_\fN\}$ are pairwise distinct,
we sometimes use the term "fermionic configuration(s)".

In our paper, the role of the graph $\cZ$ is played by the integer lattice $\DZ^d$, $d\ge 1$,
with the edges formed by the nearest-neighbor pairs $(x,y)$ (with $\|x-y\|_1=1$;
here $\|x\|_1 := \sum_i |x^{(i)}|$ for $x = (x^{(1)}, \ldots, x^{(d)})\in\DZ^d$).
We denote by $\Bx \ominus \By$
the symmetric difference of the sets $\{x_1, \ldots, x_\fN\}$
and $\{y_1, \ldots, y_\fN\}$.

Now define on $\bcZN$ the graph structure $(\bcZN, \bcEN)$ induced by the adjacency matrix
$A_{\Bx \By}(\bcZN)$ indexed by the points $\Bx$, $\By\in\bcZN$:
$$
A_{\Bx \By}(\bcZ) =
\left\{
  \begin{array}{ll}
    1, & \hbox{\;\; if  $\Bx \ominus \By\in \cE$;} \\
    0, & \hbox{\;\;otherwise.}
  \end{array}
\right.
$$
The notation $\Bx \ominus \By\in \cE$ is slightly abusive;
more formally, a pair $(\Bx,\By)$ is an edge of $\bcZN$ iff, for some $z_2, \ldots, z_\fN\in\cZ$,
$$
\Bx = (x_1, z_2, \ldots, z_\fN), \; \By = (y_1, z_2, \ldots, z_\fN),
\;
\; \#\{x_1, y_1, z_2, \ldots, z_\fN\} = \fN+1,
$$
and $(x_1, y_1)\in\cE$ is an edge of the underlying graph $\cZ$.

Pictorially, the configuration $\By$ is obtained from $\Bx$ by moving
exactly one particle (at $x_1$) to one of its nearest neighbors ($y_1$) in the
1-particle configuration space
$\cZ$, without leaving the "sector" of configurations of $\fN$ pairwise distinct positions.

The graph $(\bcZ,\bcEN)$ is known to be connected, whenever $\cZ$ is connected.

Naturally, the above definition of the adjacency can be extended to particle configurations
with duplicate positions. Since we are going to work only with anti-symmetric functions on
the Cartesian product graph $\cZ^\fN$, these functions must vanish on $\pt\bcZNge$, thus forming
a closed Hilbert subspace, corresponding to the Dirichlet boundary conditions on $\pt\bcZNge$.

The graph structure $(\bcZN,\bcEN)$ induces the canonical graph distance
$\Brho(\,\cdot\,,\,\cdot\,)$: $\Brho(\Bx,\By)$ is the length of the shortest path
$\Bx \rightsquigarrow \By$ over the edges of $\bcZN$.

The graph structure induces also the canonical graph Laplacian
$$
(-\BDelta \BPsi)(\Bx) = \sum_{\nb{\Bx,\By}\in \bcZN} (\BPsi(\Bx) - \BPsi(\By))
= \cC_{\bcZN}(\Bx) \BPsi(\Bx) - \sum_{\nb{\Bx,\By}\in \bcZN} \BPsi(\By);
$$
here  $\nb{\Bx,\By}$ denotes a pair of nearest neighbors ($\brho(\Bx,\By)=1$),
and $\cC_{\bcZN}(\Bx) $ is the coordination number (the number of nearest neighbors) of $\Bx$.

Given a subgraph $\BLam\subset\bcZN$, we define its internal, external and the
so-called graph (or edge) boundary,
in terms of the canonical graph distance:
$$
\bal
\pt^-\BLam &= \myset{ \By\in\BLam:\; \Brho(\By, \BLam^\rc) = 1}
\\
\pt^+\BLam &= \myset{ \By\in\BLam^\rc:\; \Brho(\By, \BLam) = 1} = \pt^-\BLam^\rc
\\
\pt\BLam &= \myset{ (\Bx,\By)\in\BLam\times\BLam^\rc:\; \Brho(\Bx, \By) = 1}.
\eal
$$

Given a fermionic configuration $\Bx =\{x_1, \ldots, x_\fN\}\in \bcZN$, $\fN>1$,  we call any subset
$\Bx' = \{x_j, \, j\in J\}$, with $J\subset \{1, \ldots, \fN\}$,  sub-configuration
of $\Bx$, and write $\Bx'\subset \Bx$.
Given a sub-configuration $\Bx'\subset \Bx$,
one can define the complement of $\Bx'$ relative to $\Bx$, $\Bx'' = (\Bx')^\rc := \Bx \ominus \Bx'$;
in this case, we say that $\Bx$ admits the decomposition $\Bx = (\Bx',\Bx'')$.

\subsection{Representation by occupation numbers}
\label{ssec:occup.numbers}
One can use an alternative construction of the configuration space of
$\fN$ indistinguishable (fermionic) particles. Namely, given an ordered
$\fN$-tuple $\vBx=(x_1, \ldots, x_\fN)\in\cZ^\fN$,
define a
function\footnote{In other words, $\Bn$ is a formal finite linear combination of points of
$\cZ$ with integer coefficients.}
$\Bn_{\vBx}:\cZ\mapsto \DN$
by
$
\Bn_{\vBx}(y) = \# \{j: \, x_j = y\}, \;\; y\in\cZ.
$
Then the set of vertices $\bcZ$ corresponds to the set of functions $\Bn_\Bx:\cZ\to\DZ$
with values in $\{0,1\}$ subject to the constraint $\sum_{y\in\cZ} \Bn_\Bx(y) = \fN$. In physical terms,
$\Bn_{\Bx}$ gives the "occupation numbers" of the $\fN$-particle configuration $\Bx$.
When the particles are considered indistinguishable (fermionic or bosonic),
the occupation numbers (i.e., the number of particles
of the given configuration occupying each lattice point) uniquely identify the configuration,
and vice versa. In the case considered in the present paper, where the particles are fermions,
occupation numbers take values in $\{0,1\}$. Formally, the boundary $\pt \bcZNge$ does not appear
in this alternative construction, but for the definition of the multi-particle Hamiltonians,
it can be introduced, by allowing occupation numbers $\Bn_\Bx$ with values
in $[[0, \fN]]$ subject to the constraint $\sum_{y\in\cZ} \Bn_\Bx(y) = \fN$.

\subsection{Fermionic Hamiltonians}
\label{ssec:Fermi.Hamiltonians}

A self-consistent representation of the fermionic Hamiltonian $\BH(\om;\th)$,
not referring to the lattice $(\DZ^d)^\fN$ serving as the configuration space
of the distinguishable particles, is as follows:
\be\label{eq:def.fermionic.H}
\BH = -\BDelta_{\bcZN} + g \sum_{x\in\Bx} v(T^x  \om;\th) +
\sum_{x,y \inneq \Bx} U^{(2)}(|y-x|),
\ee
where the second and the third terms in the RHS are understood as the operators
of multiplication by the functions $\Bx \mapsto g\BV(\Bx;\om;\th)$ and, respectively,
$\Bx\mapsto \BU(\Bx)$, $\Bx\in\bcZN$.

Given an $\fN$-particle DSO $\BH = \BH^{\fN}(\om;\th) = \BH_0 + g\BV(\om;\th) + \BU$,
and a proper subset $\BLam\subsetneq \bcZN$, we consider the restriction
$\BH_\BLam$ of $\BH$ to $\BLam$ defined as follows:
$\BH_\BLam = \one_\BLam \BH \one_\BLam \upharpoonright \ell^2(\BLam)$, where the indicator
function $\one_\BLam$ is identified with the multiplication operator by this function,
and also with the natural orthogonal projection from $\ell^2(\bcZN)$ onto $\ell^2(\BLam)$.
$\BH_\BLam$ is usually considered as the discrete analog of the Schr\"{o}dinger operator
with Dirichlet boundary conditions, acting on functions $\Bpsi$ vanishing outside $\BLam$.

As usual in the Multi-Scale Analysis, we will work with the length scale sequence
$\{ L_j, \, j\ge 0\}\subset\DN^*$: given a positive integer $L_0$,
we set for $j\ge 1$
\be\label{eq:scale.Lj}
L_{j} = L_{j-1}^{2} = (L_0)^{2^j}.
\ee
It is convenient for the analysis of unimodal eigenstates
to define also the  scale
\be\label{eq:scale.L.minus.one}
L_{-1} = 0,
\ee
so that the balls $\ball_{L_{-1}}(u)\subset\DZ$ and
$\bball_{L_{-1}}(\Bu)\subset\bcZN$ are single-site sets.

\subsection{Augmented dynamical system}
\label{ssec:BK.augmented.T}

The presence of $\fN$, possibly different, phase points $T^{x_1}\om$, \ldots, $T^{x_\fN}\om$
in the potential energy
$$
\BV(\Bx;\om;\th) = \sum_{i=1}^\fN v(T^{x_i}\om;\th)
$$
suggests introducing
the $\fN$-th Cartesian power of the dynamical system $T$, viz.
\be\label{eq:def.dyn.syst.BT}
\begin{array}{ccccl}
&\BT: &\bcZN\times (\DT^\nu)^\fN &\to  &(\DT^\nu)^\fN
\\
& & (x_1, \ldots, x_\fN;  \om_1, \ldots, \om_\fN) & \mapsto
           &(T^{x_1}\om_1, \ldots, T^{x_\fN}\om_\fN)
\end{array}
\ee
and the augmented hull, $\mathscr{V}: \, (\DT^\nu)^\fN \to \DR$, defined by
$$
\csV(\om_1, \ldots, \om_\fN)  = v(\om_1) + \cdots + v(\om_\fN).
$$
Naturally, we are only interested in the trajectories
of the diagonal phase points $\BomN:= (\om, \ldots, \om)$, and by slight abuse of notations,
sometimes it will be convenient to write
$$
\BV(\Bx;\BomN;\th) = \csV(\BT^\Bx \BomN), \;\;
\BH(\BomN;\th) = \BH_0 + g\csV(\BT^\Bx \BomN) + \BU,
$$
instead of our usual notations $\BV(\om;\th)$, $\BH(\om;\th)$.

Given $a\in\DZ^d$, we denote by $S^a$ the translation $S^a: x \mapsto x + a$, $x\in\DZ^d$. Further, for
$\Bx\in\bcZN$, we define the unitary shift operator $\cU^\Bx$. First, in the lattice $(\DZ^d)^\fN$
of distinguishable particle configurations, we set
$$
\cU^\Bx f(\By) := f(\By - \Bx),
$$
with $\By = (y_1, \ldots, y_\fN)$, $\Bx = (x_1, \ldots, x_\fN)\in(\DZ^d)^\fN$, and then we reduce
it to the fermionic subspace of anti-symmetric square-summable functions; the latter is already represented
as the Hilbert space $\ell^2(\bcZN)$, so we can consider $\cU^\Bx$ as a unitary operator in the space
of functions on $\bcZN$.

For any $\Bu\in\bcZN$, we have the covariance relation
$$
\BH_{\bball_L(\Bu)}(\BomN;\th) = \cU^{-\Bu}\, \BH_{\bball_L(\Bzero)}(\BT^{\Bu}\BomN;\th) \, \cU^{\Bu},
$$
which does not affect the parameter $\th\in\Th$.

Removing the external potential energy $g\BV$, we obtain the extended kinetic operator
$\BK := \BH_0 + \BU$. Unlike the full, spatially inhomogeneous Hamiltonian $\BH = \BK + g\BV(\om;\th)$,
the kinetic operator has a symmetry group, including all "diagonal" translations
$\Bx \mapsto \Bx + (a, a, \ldots, a)$, $a\in\DZ^d$. Again, the latter formula initially makes sense in
$(\DZ^d)^\fN$, but then it can be extended to $\bcZN$. Generally speaking, with an interaction of
infinite range, non-diagonal translations are not symmetries of $\BK$, but, as we will see
in Sect.~\ref{sec:equivalence.BT},
the family of symmetries of $\BK$ is much richer, whenever $\BU$ has finite range. Specifically,
considering the restrictions $\BK_{\bball_L}(\Bu)$ to the balls of (any) fixed radius $L\ge 0$,
the entire family of multi-particle operators $\{\BK_{\bball_L}(\Bu),\,\Bu\in\bcZN\}$
is decomposed into a finite number of unitary equivalence classes; moreover, this unitary equivalence
is due to properly chosen shift transformations. See the details in Sect.~\ref{sec:equivalence.BT}.
We will exploit this fact in the course of approximation of $\BH$ by truncated
Hamiltonians, with finite-range interactions.

\section{Requirements for the dynamical system and the interaction}\label{sec:UPA}

For the sake of clarity,
we consider only the case where $\Om = \DT^\nu$, $\nu\ge 1$, and it is convenient
to define the distance $\dist_\Om[\om', \om'']$ as follows: for
$\om'=(\om'_1, \ldots, \om'_\nu)$ and $\om''=(\om''_1, \ldots, \om''_\nu)$,
$$
\dist_\Om[\om', \om'']
:= \max_{1 \le i \le \nu} \dist_{\DT^1}[\om'_i, \om''_i],
$$
where $\dist_{\DT^1}$ is the conventional distance on the unit circle $\DT^1 = \DR^1/\DZ^1$.
With this definition, the diameter of a cube of side length $r$ in $\DT^\nu$ equals $r$, for any dimension $\nu\ge 1$. The reason for the choice of the phase space $\Om=\DT^\nu$ is that the parametric families of ensembles of potentials $V(x;\om;\th)$ are fairly explicit in this case, and besides, this allows one
to construct families of quasi-periodic operators.

We assume that the underlying dynamical system $T$ (generating the potential)
satisfies the condition of Uniform Power-law Aperiodicity:

\noindent
\UPA \quad $\exists\, A, C_A\in\DN^*\; \;\forall\, \om\in\Om \;
  \forall\, x, y\in\DZ^\nu \text{ such that } x\ne y $
\be\label{eq:cond.UPA}
\begin{array}{lc}
\quad \dist_{\Om}(T^x \om, T^y \om) \ge C^{-1}_A |x - y |^{-A},
\end{array}
\ee
and of tempered local divergence of trajectories:
\par
\vskip2mm

\noindent
\DIV \quad
$\exists\, A', C_{A'}\in\DN^*\; \;\forall\, \om, \om'\in\Om \;\forall\, x\in\DZ^\nu\setminus\{0\}$
\be\label{eq:cond.DIV}
\begin{array}{lc}
\quad \dist_{\Om}(T^x \om, T^x \om') \le C_{A'}\, |x|^{A'} \dist_{\Om}(\om, \om').
\end{array}
\ee

\bre
It is not difficult to see that both \UPA and \DIV rule out strongly mixing dynamical systems like
the hyperbolic toral automorphisms (while the skew shifts of tori are still allowed).
This certainly looks quite surprising, but it has to be emphasized
that our proof is oriented towards the dynamical systems with the \emph{weakest} stochasticity. In a manner
of speaking, we actually need that the dynamical system ``do not interfere'' with the ``randomness''
provided by the parametric freedom in the choice of the sample potential $V(\cdot;\om;\th)$. As to
the mixing systems, their intrinsic stochasticity is to be used in the proof of localization
in a different way; this puts them beyond the scope of the present paper. Note, however, that
the localization properties of deterministic DSO with strongly mixing potential should, in our opinion,
be similar to those of the genuinely random DSO. In particular, we believe that
the uniform decay and unimodality of the eigenfunctions cannot occur for the DSO with sufficiently
strongly mixing potentials.
\ere

For the rotations of the torus $\DT^\nu$,
\DIV holds trivially, since $T^x$ are isometries,
and \UPA reads as the Diophantine condition for the frequencies.

\vskip1mm
Finally, we make the following assumption:

\vskip1mm
\noindent
\Uone The two-body interaction potential $U$ satisfies the decay bound
\be\label{eq:decay.U}
\forall\, r \ge 1\quad |U(r)| \le  r^{- 2B \ln r},
\ee
where $B = 400 b A^2/\ln 2$  (cf. \eqref{eq:def.B}) with the same value $A>0$ as in \UPA
and $b>0$ as in \eqref{eq:def.an.b}.

\section{Parametric ensembles of potentials}
\label{sec:param.ensembles}

\subsection{The Local Variation Bound}
\label{ssec:LVB}

Following \cite{C12a}, we introduce now a hypothesis on the random field $v:\Om\times\Th\to\DR$
on $\Om$,
relative to the probability space  $(\Th,\fB,\prthp)$, which is logically independent of the
particular construction given in subsection \ref{ssec:randelettes}. Later we will show that it is
holds true for the hull functions $v$ constructed with the help of the randelette expansions described in
subsection \ref{ssec:randelettes}.

\vskip2mm
\LVB:
\emph{
Let $v:\Om\times\Th\to\DR$ be a measurable function on the product probability space
$(\Om\times\Th, \fF\times\fB,\DP\times\prthp)$.
There exists a family of sub-sigma-algebras $\fB_L \subset \fB$, $L\in\DN^*$, such that, conditional
on $\fF\times\fB_L$ (hence, with $\om\in\Om$ fixed), for any ball $\ball_{L^4}(u)$, the values
$\{V(x;\om;\th):=v(T^x \om;\th), \, x\in \ball_{L^4}(u)\}$, are (conditionally) independent and admit individual (conditional) probability densities $\rhoL_{v,x}(\cdot\,|\fF\times\fB_L)$, satisfying,
for some $B, C''\in(0,+\infty)$,}
\be\label{eq:LVB}
\essup
\big\| \rhoL_{v,x}(\cdot\,|\fF\times\fB_L) \big\|_\infty \le C'' L^{B \ln L}.
\ee
\vskip1mm

In our model, the key probability estimates will be established for all $\om\in\Om=\DT^\nu$.

\vskip1mm
It is readily seen that for the scaled random variables $(\om;\th) \mapsto g V(x;\om;\th)$
the assumption \eqref{eq:LVB} implies
\be
\essup
\| \rhoL_{gv,x}(\cdot\,|\fF\times\fB_L) \|_\infty
\le C'' g^{-1}  L^{B \ln L}, \;\; C''\in(0,+\infty).
\ee

This property allows to prove analogs of the Wegner estimate (cf. Sect. \ref{ssec:Wegner})
in finite balls of any size $L$; in our model, such estimates, as we shall see,
are not uniform in $L$ and actually deteriorate as $L\to\infty$.

As shows the proof of Lemma  \ref{lem:LVB}, for the class of models considered in this paper, the above
mentioned individual densities $\rhoL_{v,x}$ are simply uniform densities, in some intervals
$I^{(L)}_{v,x}\subset\DR$ of length $a_{\tN(L)}$, thus bounded by
$a^{-1}_{\tN(L)} \le \Const L^{O(\ln L)}$.

\subsection{Regularity of the Conditional Mean}
\label{ssec:RCM}

The additional assumption formulated below is not critical for the proof of the
$\fN$-particle \emph{semi-uniform} localization in the entire lattice $\bcZN$
(which is beyond the scope of the present paper).
However, it is required for the efficient \textbf{multi-particle} localization bounds in
(arbitrarily large) finite volumes. This feature
is not specific to the deterministic operators; in fact, it was first observed
in the context of $\fN$-particle DSO with IID random potential (cf. \cite{CS09b}, \cite{AW09a}).
A partial solution to this technical problem was proposed in our earlier work
\cite{C11a}, for a class of random potentials including Gaussian random fields
with discrete or continuous argument, and later extended to
a larger class of random potentials.
Here we adapt it to the context of deterministic potentials.

\vskip2mm
\RCM:
\emph{
Let $v:\Om\times\Th\to\DR$ be a measurable function on the product probability space
$(\Om\times\Th, \fF\times\fB,\DP\times\prthp)$. There exist constants
$a', b', C'\in(0,+\infty$ with the following property.
\vskip1mm
Let $\Lam\subset\DZ^d$ be a finite set, and consider the empirical mean $\xi_\Lam(\om;\th)$
of the sample $\{V(x;\om;\th):=v(T^x \om;\th), \, x\in \Lam \}$ and the fluctuations relative to
$\xi_\Lam(\om;\th)$:
}
\be\label{eq:eta.Lam}
\xi_\Lam(\om;\th) := \frac{1}{|\Lam|} \sum_{x\in\Lam} V(x;\om;\th),
\quad \eta^{\Lam}_x(\om;\th) := V(x;\om;\th) - \xi_\Lam(\om;\th),
\;\; x\in\Lam.
\ee
\emph{Next, let $\fB^\Lam_\eta$ be sub-sigma-algebra of $\fF\times\fB$ generated by the random
variables
\be\label{eq:fB.eta.Lam}
\{\eta_x^\Lam(\cdot), \, x\in\Lam; \;V(y;\cdot;\cdot), \, y\in \cZ \setminus \Lam   \}.
\ee
Then for any $\fF\times\fB^\Lam_\eta$-measurable function $\mu:\Om\times\Th\to\DR$,
one has
\be\label{eq:RCM}
\forall\, s\in(0,1]\;\; \essup \; \prth{ \xi_\Lam \in[\mu, \mu + s] \cond \fB_\eta \;  } \le C'' |\Lam|^{b''} s^{a''}.
\ee
}
\vskip2mm

\noindent
In our work \cite{C13a}, the property \RCM was proven for the uniform (and some other)
probability distributions. See Proposition \ref{prop:RCM.Unif} in Appendix
\ref{sec:App.RCM.Wegner}, where it is used in the proof of the key Wegner-type estimate,
Theorem \ref{thm:weak.sep.Wegner}.

\subsection{Deterministic  potentials and randelette expansions}
\label{ssec:randelettes}

In Ref. \cite{C11c}, we introduced \textit{parametric families} of ergodic ensembles of operators
$\{H(\om;\th), \om\in\Om\}$ depending upon a parameter $\th\in\Th$ in an auxiliary space $\Th$.
As shows \cite{C11c}, it is convenient to endow $\Th$ with the structure of a probability
space, $(\Th, \fB, \prthp)$, in such a way that $\th$ be, in fact, an \textit{infinite} family of
IID random variables on $\Th$, providing an infinite number of auxiliary independent
parameters allowing to vary the
hull $v(\om;\th)$ locally in the phase space $\Om$. We called such parametric
families \emph{grand ensembles}.

The above description is, of course, too abstract. In the framework of the DSO,
we proposed in  \cite{C11c} a more specific construction where $H(\om;\th) = H_0 + V(\cdot;\om;\th)$,
with $V(x;\om;\th) = V(T^x\om;\th)$ and
\be\label{eq:randelettes.1}
v(\om; \th) = \sum_{n=1}^\infty a_n \sum_{k=1}^{K_n} \th_{n,k} \ffi_{n,k}(\om),
\ee
where the family of random variables $\th := (\thnk, n\ge 1, 1 \le k \le K_n)$ on $\Th$ is IID, and
$\ffink := (\ffink), n\ge 1, 1 \le k \le K_n<\infty)$ are some functions on the phase space $\Om$ of the
underlying dynamical system $T^x$, such that
$\diam \supp\, \ffink \to 0$ as $n\to\infty$. Representations of the form \eqref{eq:randelettes.1}
were called in \cite{C11c} \textit{randelette}  expansions, referring to the "random" nature of the
expansion coefficients and to the shape of $\ffi_{n,k}$ reminding of the wavelets ("ondelettes", in French).

Putting the amplitude of the function $\ffink$ essentially
in the "generation" coefficient $a_n$, it is natural to assume
that $|\ffink(\om)|$ are uniformly bounded in $(n,k,\om)$.
Further, in order to control the potential  $V(T^x \om;\th)$ at any lattice site $x\in\DZ^d$
or, equivalently, at every point $\om\in\Om$, it is natural to require that for every $n\ge 1$,  $\Om$
be covered by the union of the sets where at least one function $\ffink$ is nonzero (and preferably, not
too small).

In the Sect.~\ref{ssec:lac.haar.rand} below, we make a specific choice for $\{a_n\}$ and $\{\ffink\}$.

Notice that the dynamics $T^x$ leaves $\th$ invariant, so the latter is merely a parameter
in the deterministic potential $V(x;\om;\th) = v(T^x\om; \th)$, generated by the values of the hull function
$v_\th(\cdot): \om \mapsto v(\om;\th)$ measured along the trajectory $\{T^x\om, \, x\in\DZ^d\}$
of the dynamical system $T:\DZ^d\times \Om\to\Om$.

\subsection{Lacunary ``haarsh'' randelette expansions}
\label{ssec:lac.haar.rand}

In the present paper, we focus on the case is where the randelettes are piecewise constant Haar wavelets.
For example, if $\Om = \DT^1 = \DR/\DZ$, we set, for
$n\ge 0$, $1 \le k \le K_n = 2^n$,
$$
\ffink(\om) = \one_{\Cnk}(\om), \;\; \Cnk = \left[(k-1)2^{-n}, k2^{-n}\right) .
$$
It is to be emphasized that the orthogonality of the system of functions
$\{\ffink, n\ge 0, 1\le k \le K_n\}$
is not essential for our results and proofs. In fact, one could take the functions
$|\ffink|$ (which are obviously non-orthogonal); this would even result in slightly simpler
arguments in some auxiliary statements. Such a model was actually studied in \cite{C11c}.

On the torus of higher dimension, one has to replace intervals of length $2^{-n}$ by cubes of sidelength $2^{-n}$.
Specifically, given an integer $n\ge 0$, for each integer vector $(l_1, \ldots, l_\nu)$ with $1 \le l_i \le 2^n$, consider the cube
$$
\mytimes_{i=1}^{\nu} \left[(l_i-1)2^{-n}, l_i 2^{-n}\right)
\subset \DT^\nu.
$$
These cubes can be numbered in some way,
e.g., in the lexicographical order of vectors $(l_1, \ldots, l_\nu)$;
their  number equals
$K_n = 2^{\nu n}$. We will denote these cubes by $C_{n,k}$, $k=1, \ldots, K_n$.
Next, introduce a countable family of functions on $\DT^\nu$,
$$
\ffink(\om) = \one_{C_{n,k}}(\om), \;n\ge 0, \;k=1, \ldots, K_n,
$$
and a  family of IID random variables $\thnk$ relative to
an auxiliary probability space $(\Th,\fB,\prthp)$, uniformly distributed in $[0,1]$.
Next, pick a number $b > 2\, \fN\, d$ and set
\be\label{eq:def.an.b}
a_n = 2^{-2bn^2}, \; n\ge 1.
\ee
Now define a function $v(\om;\th)$ on $\Om \times \Th$,
\be\label{eq:hull.haarsch}
v(\om;\th)  = \sum_{n=1}^\infty a_n \sum_{k=1}^{K_n} \th_{n,k} \ffi_{n,k}(\om),
\ee
which can be viewed as a family of functions $v(\cdott;\th)$ on the torus, parameterized by $\th\in\Th$,
or as a particular case of a "random" series of functions, expanded over a given system of functions
$\ffink$ with "random" coefficients.

We will call the expansions of the form \eqref{eq:hull.haarsch}
\emph{haarsh},  referring to Haar's (\emph{Haarsche}, in German) wavelets and to the "harsh" nature of the
resulting potentials. Constructing a potential out of flat pieces is rather unusual in the framework of
the localization theory, where
all efforts  were usually made so as to avoid flatness of the potential.
Yet, with an infinite number of flat components $\thnk\,\ffink(\om)$, each modulated by its own
parameter $\thnk$, we proved earlier (cf. \cite{C07a,C11c,C13a})
an analog of Wegner bound \cite{W81} for
the respective grand ensembles $H(\om;\th)$.

The extremely rapid decay of the coefficients $a_n$, making the series "lacunary", is required
for the proof of unimodality and of uniform decay of eigenfunctions. With coefficients behaving like
$a_n \sim 2^{-bn}$, the sum of the tail series $\eps_{N+1}:=\sum_{n\ge N+1} a_n$ is comparable to $a_N$, while
we need $\eps_{N+1} \ll a_N$.

Building on the techniques from \cite{C11c,C13a,C13b}, we prove Anderson localization for
generic lacunary haarsh  potentials of sufficiently large amplitude,
under the mild assumptions \UPA (cf. \eqref{eq:cond.UPA}) and \DIV (cf. \eqref{eq:cond.DIV}).
In particular, we prove Anderson localization for a class of
quasi-periodic potentials with Diophantine frequencies.
As in \cite{C11c,C13b}, we use a variant of the Multi-Scale Analysis, developed in \cite{FMSS85,DK89}
for random operators.

\section{Main results}
\label{sec:Main.res}

\btm\label{thm:Main}
Let be given arbitrary integers $\fN, d \ge 1$, and consider the $\fN$-th symmetric
power $\bcZN$ of the lattice $\DZ^d$, with the conventional graph structure.
Consider a family of fermionic $\fN$-particle Hamiltonians in $\ell^2(\bcZN)$,
of the form
\eqref{eq:def.fermionic.H}
Assume that  the external potential has the form
$V(x;\om;\th) = v(T^x\om;\th)$ with $v(\om;\th)$ given by the expansion \eqref{eq:randelettes.1}, and the
dynamical system $T^x$ satisfies conditions \UPA and \DIV  (cf. \eqref{eq:cond.UPA}, \eqref{eq:cond.DIV})
for some $A, C_A, A', C_{A'} \in\DN^*$. Finally, assume that the interaction potential satisfies
the hypothesis {\rm\Uone}.

Then there exist finite positive constants $g_0$,  $C_0$, and $c'$,
depending upon the parameters $A, C_A, A', C_{A'},d, \fN, \nu$,
such that for any $g\ge  g_0$,  there exists a subset
$\Thinf(g) \subset \Th$ of measure
$$
\prth{\Thinf(g)} \ge 1 - C_0\, \beta(g),
$$
where
$$
0 < \beta(g) \le \eu^{-c' \ln^{1/2} g},
$$
with the following property: if $\th\in\Thinf(g)$, then for \textbf{any} $\om\in\Om$:
\begin{enumerate}[\rm(A)]
  \item $\BH(\om;\th)$ has pure point spectrum;

  \item for any $\Bx\in\bcZN$, there is exactly one eigenfunction $\Bpsi_{\Bx}(\cdott;\om;\th)$ such that
\be\label{eq:psi.loc.center}
|\Bpsi_{\Bx}(\Bx;\om;\th)|^2 > 1/2,
\ee
  i.e., $\Bpsi_{\Bx}$ has the ``localization center'' $\Bx$, and the localization centers
  establish a bijection between the eigenbasis $\{\Bpsi_{\Bx}(\cdot;\om;\th)\}$
  and the graph $\bcZN$;

  \item for all $\Bx\in\bcZN$, the eigenfunctions $\Bpsi_{\Bx}$ decay uniformly exponentially fast
  away from their respective localization centers:
$$
\forall\,  \By\in\bcZN\;\; |\Bpsi_{\Bx}(\By;\om;\th)| \le  \eu^{-m \rd(\By,\Bx)}, \; m = m(g, C, A)>0.
$$
\end{enumerate}
\etm

The proof of Theorem \ref{thm:Main}, or rather the last, mainly analytic part of this proof, with the help
of a scale induction, occupies Sections \ref{sec:MSA}--\ref{sec:Uniform};
it relies upon the eigenvalue concentration
estimates established in Sections
\ref{sec:randelettes.and.sep.bounds}--\ref{sec:sep.spec.L0} and \ref{sec:proof.key.tech.results}
for all length scales, without the scale induction.

\vskip3mm
We would like to emphasize, especially for the readers familiar with the works
\cite{CS09b,CBS11} and the recent works by Klein and Nguyen \cite{KN13a,KN13b}
on multi-particle disordered systems, that the proof of Theorem  \ref{thm:Main}
\emph{does not use the induction on the number of particles}. However, this new technique
does not allow one to prove the uniform $\fN$-particle localization bounds
in a deterministic potential
for a system of $\fN \sim \rho |\Lam|$ particles in an arbitrarily large volume $\Lam\subset\DZ^d$,
with a fixed $\rho>0$.

\btm\label{thm:DL.uniform}
Under the assumptions and with notations of Theorem \ref{thm:Main},
let $\th\in\Thinf(g)$. If $m = m(g)>0$ is large enough, then for any $\om\in\Om$ and all
$\Bx,\By\in\bcZN$,
for any continuous function $\phi$ with $\|\phi\|_\infty\le 1$
and some constant $\Const(\fN, d)$ independent of $\om$,
$$
\sup_{\phi\in\csB_1(\DR)}
|\langle \one_\Bx | \phi(H(\om;\th)) | \one_\By \rangle|
\le \Const(\fN, d)\, \eu^{-m \Brho(\Bx, \By)}.
$$
In particular,
$$
\sup_{\om\in\Om} \;
\sup_{t\in\DR}
|\langle \one_\Bx | \eu^{ - \ii t \BH(\om;\th)} | \one_\By \rangle|
\le \Const(\fN, d)\, \eu^{-m \Brho(\Bx, \By)}.
$$
\etm

Here $\csB_1(\DR)$ stands for the set of continuous functions $\phi:\DR\to\DC$ with $\|\phi \|\le 1$.

The derivation of the uniform dynamical localization
from the results on uniform decay of unimodal eigenstates is quite simple and does not
require the usual techniques of the multi-scale analysis (cf. \cite{GDB98,DS01,GK01,GK13}).
See the proof in Sect. \ref{sec:DL.uniform}.

\section{Randelettes and separation bounds for the potential}
\label{sec:randelettes.and.sep.bounds}

\subsection{Partitions}
\label{ssec:partitions}

Consider the phase space $\Om$ which we always assume in this paper to be the torus
$\DT^\nu$ of dimension $\nu \ge 1$: $\DT^\nu = \DR^\nu/\DZ^\nu \cong [0,1)^\nu$.
For each $n\ge 0$, we have introduced the family of
$K_n = 2^{\nu n}$ adjacent cubes $\Cnk$, $k=1, \ldots, K_n$,
of side length $2^{-n}$, and functions $\ffink:\om \mapsto \one_{\Cnk}(\om)$.
More precisely, we assume that $C_{n,k}$ have the form
$$
C_{n,k} = \mytimes_{i=1}^\nu \left[l_{i,n,k} 2^{-n}, (l_{i,n,k} + 1) 2^{-n} \right),
\;\; l_{i,n,k} \in [[0, 2^n - 1]], \; i=1, \ldots, \nu,
$$
so for every $n\ge 0$, the supports $\{\Cnk = \supp \ffi_{n,k}$,  $1 \le k \le K_n \}$
naturally define a partition of the phase space $\Om$:
$$
\cC_n = \myset{ C_{n,k}, 1\le k \le K_n  }.
$$

To each element $C_{n,k}$ of the partition $\cC_n$ corresponds a unique finite sequence of indices
$\kappa(n,k) =(k_0, \ldots, k_{n-1}, k_n=k)$ labeling $n+1$ elements
$C_{i,\hk_i} \supset C_{n,k}$, $0 \le i \le n$, of partitions preceding or equal to $\cC_n$.
Here we denoted by $\hk_i(\om)$ the (unique) index such that
\be\label{eq:hk}
\om \in C_{i,\hk_i(\om)}.
\ee

\subsection{Piecewise-constant approximants of the hull}
\label{ssec:piecewise.approx}

For each $N\ge 0$, introduce the approximant of   $v(\om;\th)$ given by \eqref{eq:randelettes.1}:
\be\label{eq:v.xi.n.k}
\bal
v_N(\om;\th) &= \sum_{n=0}^N a_n \sum_{k=1}^{K_n} \th_{n,k} \ffi_{n,k} .
\eal
\ee
The random variables $\th \mapsto v_N(\om;\th)$ are strongly correlated via the values $\th_{n',\bullet}$
with $n'<n$.
However, for any fixed $n$, the family of random variables on the probability
space $(\Th,\fB,\prthp)$,
$\{\thnk(\th)(\om), \, k=1, \ldots, K_n\}$, is independent.
We shall see that the amplitudes $\thnk$ bring enough "innovation"
into the $n$-th generation of the randelettes, and imitate, to a certain degree, some important
properties of IID random potentials.

Further,  for any $N\ge 1$, if $b \ge 2$, then we have
\be\label{eq:aN.decay.fast}
\bal
\sum_{n=N+1}^\infty a_n &= \sum_{n=N+1}^\infty 2^{-bn^2}
= 2^{-b(2N+1)} 2^{-bN^2} \sum_{i=0}^\infty 2^{-b(N+i)^2 + b(N+1)^2}
\\
&\le 2^{-b(2N+1)} a_N \sum_{i=0}^\infty 2^{-i} \le \half  2^{-2bN} a_N,
\eal
\ee
so the norm
$\| v - v_N \|_\infty :=  \sup_{\om\in\Om} \|  v - v_N \|_{L^\infty(\Th)}$
can be bounded as follows:
\be\label{eq:norm.v.vN}
\|  v - v_N \|_\infty \le
\half 2^{-2bN} a_N.
\ee
Owing to \eqref{eq:aN.decay.fast}, the RHS is \textit{much smaller} than the width ($a_N$)
of the distribution of the random coefficients $a_N\th_{N,k}$,
$1 \le k \le K_N$ (recall that $\th_{N,k} \sim Unif[0,1]$).
Set
\be\label{nt.A.C}
\bal
\tn(L) &= \tn(L,A,C) :=  1 + \left\lfloor \frac{4A \ln L - \ln (C/2)}{ \ln 2} \right\rfloor ,
\\
 \tN(L) &= \tn(L^4),
\eal
\ee
and observe that, for $L$ large enough so $|\ln C|+2\ln 2 < A \ln L$,
\begin{spreadlines}{0.9em}
\be\label{eq:nt.A.C.less}
\bal
\diy
 \frac{3 A \ln L}{ \ln 2} & < \;\tn(L) & < &\; \diy\frac{5A \ln L}{ \ln 2},
\\
\diy
  L^{-20 A } &<  \;2^{-\tN(L)} & < & \;\;L^{- 12 A } .
\eal
\ee
\end{spreadlines}
Therefore,
\be\label{eq:def.tA}
2^{-\tN(L)} = L^{-\tA}, \;\;
   \tA=\tA(A,C_A) \in \left[ \frac{12 A}{\ln 2}, \, \frac{20 A}{\ln 2} \right]
\subset \left[ 17 A, \, 29 A \right].
\ee
The condition $A \ln L_0 > |\ln C|+2\ln 2$, along with some other lower bounds on $L_0>1$,
will be always assumed below, in the course of the scale induction,
where we work with balls of side length $L_j = (L_0)^{2^j}$, $j\ge 0$ (cf. \eqref{eq:Lj}).
Then for any $u\in\DZ^d$ and any $\om\in\Om$, all the points of the finite trajectory
$\{T^x\om, x\in\ball_{L^4}(u)\}$ are separated by the elements of the partition $\cC_{\tNL}$, since by
\UPA and
the first LHS inequality in \eqref{eq:nt.A.C.less}, we have
\be\label{eq:TNL.dist}
\half \dist_{\Om}(T^x\om, T^y\om) \ge \half C(L^4)^{-A} >  2^{-\tNL}.
\ee

\ble\label{lem:LVB}
Under the assumptions \UPA and \DIV,
the bound \LVB holds true with $C''=1$ and $B = 400 b A^2/\ln 2$:
\be\label{eq:lemma.ell.N.lower.bound}
a^{-1}_{\tNL} \le  L^{- B \ln L}.
\ee
\ele

\proof
Fix any integer $L\ge 1$ and let $\fB_L$ be the sigma-algebra generated by the random variables
$\{\thnk, n\le \tNL, 1\le k \le K_n\}$. By \eqref{eq:TNL.dist},
all the points of the finite trajectory $\{T^x\om, x\in\ball_{L^4}(u)\}$ are separated by the
elements of the partition $\cC_{\tNL}$, so each value $v(T^x\om;\th)$ has the form
$$
\bal
v(T^x\om;\th) &= \sum_{n < \tNL} \sum_{k=1}^{K_n} a_n \thnk \ffi_{n,k}(T^x\om)
+ \sum_{n \ge\tNL} \sum_{k=1}^{K_n} a_n \thnk \ffi_{n,k}(T^x\om)
\\
& = \zeta_\om(\th) + a_\tN \th_{\tN,\hk(T^x\om)} \fs_{\tN,\hk_\tN}(T^x\om),
\;\; \fs_{\tN,\hk_\tN}(T^x\om)\in\{-1,+1\},
\eal
$$
where the first term $\zeta_\om(\th)$ in the RHS
is $\fB_L$-measurable.
Recall (cf. \eqref{eq:hk}) that for each fixed $\om'\in\Om$,
we introduced the indices $\hk_n(\om')$ such that $\om'\in C_{n,\hk_n(\om')}$.
Since
$\thnk\sim\Unif([0,1])$ and $\fs_{\tN,\hk_\tN}(\om) = \pm 1$,
the last term in the above RHS, as a random variable on $(\Th,\prthp)$,
has probability density bounded by
\be\label{eq:sum.a.tN}
a_{\tN}^{-1} = 2^{b \tN^2} \le
\exp\left\{  \ln 2\cdot b \; \frac{ (20 A)^2 \ln^2 L}{ \ln^2 2} \right\}
=  L^{B \ln L}
\ee
with
\be\label{eq:def.B}
B = 400\, b \,A^2 / \ln 2,
\ee
and that last term is independent of $\fB_L$.
This proves the claim.
\qedhere

\section{ Wegner-type bounds and spectral spacings}\label{WS_DoS}

Recall that we have introduced in \eqref{eq:scale.Lj}--\eqref{eq:scale.L.minus.one} a sequence of integers
(length scales) $L_j, j\ge -1$, where $L_{-1} = 0$ and for all $j\ge 1$,
\be\label{eq:Lj}
L_{j} :=  L_{j-1}^{2} = (L_0)^{2^j}.
\ee

In the scale induction carried out in Sect. \ref{sec:MSA}, we will use two sequences of positive real numbers,
\be\label{eq:def.delta.j}
\bal
\delta_j  &= \delta_j = \beta_j(g) a_{\tN(L_j)},
\\
\beta_j  &= \beta_j = 2^{ - 2 b\tN(L_j)}.
\eal
\ee
It is not difficult to see that
\be\label{eq:beta.g}
\beta(g) \le \eu^{ - c_2 \ln^{1/2} g},
\ee
with some $c_2>0$ which will be specified later.
Owing to \eqref{eq:nt.A.C.less}, we have
\be\label{eq:bounds.on.delta.j}
\bal
\delta_j
&< C_1 \, L_j^{- (3A)^2 b \cdot 4\ln L_j}
\le C_2\, L_0^{- C' 2^j \ln L_0},
\eal
\ee
so that
$$
S(g) := \sum_{j\ge 0} \delta_j
< \infty, \quad
S(g) \tto{g\to\infty} 0.
$$

\subsection{Weak separation of balls}

In the conventional, single-particle localization theory of discrete Schr\"{o}dinger operators
with IID random potential, the operators $H_\Lam(\om)$, $H_{\Lam'}(\om)$ with
$\Lam\cap\Lam'=\varnothing$ are independent; one of the most important distinctions of the
multi-particle localization analysis is that the dependence between the operators
$\BH_{\bball_L(\Bx)}(\om)$ and $\BH_{\bball_L(\By)}(\om)$ does not necessarily vanish,
nor decays in any way, as $\|\Bx - \By\|\to\infty$. A reasonable replacement for the property,
usually called \emph{Independence At Distance} (IAD) in the single-particle theory, was proposed
in our works \cite{CS09a,CS09b}; we called it there "separability" of multi-particle balls.
However, the "separability" introduced in \cite{CS09a,CS09b} was too restrictive; specifically,
it is not necessarily satisfied by distant $\fN$-particle balls. A more general
notion, \emph{weak separability}, was proposed in \cite{C10}, to address the problem of efficient multi-particle
EVC bounds. In the present paper, where the particles are assumed to be indistinguishable,
the notion of weak separability from \cite{C10}
is to be adapted to the context of the $\fN$-th symmetric power of $\DZ^d$.

\bde
Let $L\ge 0$ and $\bballn_L(\Bx)$, $\bballn_L(\By)$ be two $\fn$-particle
balls, $\fn\ge 2$.
We say that $\bballn_L(\Bx)$ is weakly separated from $\bballn_L(\By)$ if there exist:
\begin{enumerate}
  \item a decomposition $\Bx = (\Bx',\Bx'')$, $\Bx' \in\bcZ^{\fn_1} $, $1 \le \fn_1\le \fn$,

  \item a similar decomposition
  $\By = (\By',\By'')$, $\By' \in\bcZ^{\fn_2} $, $0 \le \fn_2 < \fn_1$,

  \item a cube $Q \subset\DZ^d$   with $\diam Q \le 2 \fn L$,
\end{enumerate}
such that
\be\label{eq:def.weak.sep}
\bal
\Pi \Bx' \cup \Pi \By' &\subset Q,
\\
\Pi \Bx'' \cup \Pi \By'' &\subset \DZ^d \setminus Q .
\eal
\ee
A pair of balls $\bballn_L(\Bx)$, $\bballn_L(\By)$ is called weakly  separated if at least
one of the balls is weakly separated from the other.
\ede

Given a subset $\BLam\subset \bcZN$,
sometimes we will write $\bball_L(\Bx), \bball_L(\By)\sqsubset \BLam$, meaning
that $\bball_L(\Bx), \bball_L(\By)\subset \BLam$ and the balls $\bball_L(\Bx), \bball_L(\By)$
are weakly separated.

To refer explicitly to the set $Q$ figuring in the property \eqref{eq:def.weak.sep}, where appropriate,
we will say that the pair of balls $\bball_L(\Bx)$, $\bball_L(\By)$ satisfying \eqref{eq:def.weak.sep}
is weakly  $Q$-separated.
The following elementary statement shows that "sufficiently distant" balls are weakly separated.

\ble[Cf. \cite{C10}*{Lemma 2.3}]
\label{lem:3NL.WS}
A pair of balls $\bballn_L(\Bx)$, $\bballn_L(\Bx)$ with $\brho(\Bx,\By) > 3 \fn L$ is weakly separated.
In particular, distinct single-point subsets $\{\Bx\}\equiv \bballn_0(\Bx)$ and
$\{\By\}\equiv \bballn_0(\By)$, $\Bx\ne\By$, are weakly separated.
\ele
\proof
See the proof in Appendix \ref{sec:proof.Lemma.lem:3NL.WS}.
\qedhere

\subsection{Wegner-type bounds}
\label{ssec:Wegner}

Given an $\fN$-particle ball $\bballN_L(\Bu)$,
we will
denote by $\Sigma\big(\bballN_L(\Bu)\big)$ the set of eigenvalues $E_i$ (counting multiplicities)
of the respective Hamiltonian $\BH_{\bballN_L(\Bu)}(\om;\th)$;
they will be assumed to be numbered in non-decreasing order,
so that $E_i \le E_{i+1}$.
The distances $|E_{i+1} - E_i| $ will be called the spectral spacings (for  $H$).
Given the spectra of two operators, $\{E'_i\}$ and $\{E''_j\}$,
we sometimes refer to the distances $|E'_i - E''_j|$ as \emph{inter-spectral} spacings.

The following theorem is an adaptation of the main result of \cite{C10}.

\btm
\label{thm:weak.sep.Wegner}
Suppose that the grand ensemble, generated by the randelette expansion of the form
\eqref{eq:randelettes.1},
satisfies the hypotheses \UPA, \DIV and \LVB.
Then for any $j\ge 0$ and any
$3 \fN {L_j}$-distant pair of balls $\bballN_{L_j}(\Bx)$, $\bballN_{L_j}(\By)$ one has
\be\label{eq:Thm.Wegner.WS}
\bal
\pr{ \dist\left( \Sigma\left(\bballN_{L_j}(\Bx) \right),
    \Sigma\left(\bballN_{L_j}(\By) \right) \right) \le g s }
&\le  C_5 {L_j}^{(2\fN +4)d +B \ln {L_j}} \, s^{2/3},
\eal
\ee
with $C_5 = C_5(\fN,d)\in(0,+\infty)$.
\etm

\proof See the proof in Appendix \ref{sec:App.RCM.Wegner}.
\qedhere

\bre
The estimate \eqref{eq:Thm.Wegner.WS} is unusual in two ways:

$\bullet$
As in \cite{C11c}, and unlike many works on random Anderson-type Hamiltonians,
it applies to the operators with deterministic potentials.

$\bullet$ Unlike earlier published papers on multi-particle Anderson localization, it applies
to any pairs of $O(L)$-distant balls of radius $L$, with the distance induced by a
conventional norm-distance on the lattice. In \cite{CS09b,AW09a,KN13a,KN13b}, the so-called
Hausdorff distance has been used, explicitly or implicitly. This did not allow one to
establish satisfactory Wegner-type bounds in arbitrarily large, but finite volumes,
albeit one still could prove localization in the infinite lattice. Such a situation
was hardly acceptable for the applications to physical models, where a sample of a disordered media
has always a finite size.

\ere

\section{Separation bounds for the local spectra}
\label{sec:sep.spec.L0}

\subsection{Initial scale}

Given a ball $\bball:=\bball_{L_0}(\Bu)$ and a function $\BV:\bball\to\DR$, we set
\be\label{eq:def.sep.V}
\Sepb{V, \bball} := \min\big\{ \,|\BV(\Bx) - \BV(\By)|,\; \Bx,\By\in\bball, \, \Bx\ne \By \,\big\},
\ee
(here "Sep" stands for "separation [bound]"). Observe that the eigenvalues of the operator of
multiplication by $\BV$, acting in $\ell^2(\bball)$, are precisely the values of the function
$\BV$ taken on $\bball$.

Similarly, let $\{E_j, j=1, \ldots, |\bball|\} = \Sigma\big(\BH_\bball\big)$;
then we set
\be\label{eq:def.sep.H}
\Sepb{ \Sigma(\BH_\bball) } :=
\min\big\{ \,|E_i - E_j|,\; E_i, E_j\in\Sigma(\BH_\bball), \, i\ne j \, \big\} .
\ee

\btm\label{thm:Th.bound.Sep.H.L0}
For any $m\ge 1$ there exists
an integer $L_0(g) \ge \eu^{c_1 \ln^{1/2}g}$ and positive numbers  $g_*(m)>0$,
$\beta_0(g) \le \eu^{-c_2 \ln^{1/2} g}$
and a subset $\Th^{(-1)}(g)\subset\Th$ of measure
$$
\prth{\Th^{(-1)}(g)} \ge 1 - C'' L_0^{- 12 b A +A''} \beta_0(g),
$$
with
$$
A'' :=  8\fN d + 4\nu(A + A'), \;\; C'' = C''(A, A', \fN, L_0),
$$
such that for all $(\om,\th)\in\Om\times\Th^{(-1)}(g)$ and any $\Bu$,
one has
\be\label{eq:def.Th0.sep.V.firsttime}
\Sep{ g\BV(\cdot\,;\om;\th), \bball_{L_0^4}(\Bu) } \ge 4 g \delta_0 \ge 16\, \fN d \eu^{4m} .
\ee
\etm

See the proof in Sect.~\ref{ssec:entropy.bounds}.

\subsection{Arbitrary scale}

Introduce the following quantities,
\begin{align}\label{eq:D.L.om.th.1}
D(L, \om, \theta; \Bx, \By)
& = \dist\left(\Sgomth{L}{\Bx}, \; \Sgomth{L}{\By} \right) \\
\label{eq:D.L.om.th.2}
D(L, \om, \theta)
& =
\min_{ \lr{\bball_L(\Bx), \,\bball_L(\By)} \sqsubset \bball_{L^{4}}(\Bzero) }
\;\;
D(L, \om, \theta; \Bx, \By), \;\;
\\
\label{eq:D.L.om.th.3}
D(L, \theta)
& =  \inf_{\om\in\Om} \; D(L, \om, \theta),
\end{align}
and the sets
\be\label{eq:def.Thinf}
\bal
\Th^{(j)}(g) &:= \myset{\th\in\Th:\,  \inf_{\om\in\Om} D(L_j,\th) \ge 4g \delta_j },
\\
\Thinf(g) &:= \cap_{j\ge 0} \Th^{(j)}(g).
\eal
\ee

\btm\label{thm:cor.thm.weak.DLom}
Suppose that the conditions \UPA and \DIV are fulfilled,
with $b > A^{-1}\big(16\fN d + 8\nu(A + A')\big)$. Then for $L_0$ large enough
\be
\prth{ \Th^{(j)}(g)} = \prTh{ \inf_{\om\in\Om} D(L_j,\th) \ge 4 g \delta_j} \ge 1 - L_j^{- bA },
\ee
and, therefore,
\be\label{eq:mu.Thinf}
\prth{ \Thinf(g)}  \tto{g\to\infty} 1 .
\ee
\etm

See the proof in Sect.~\ref{thm:cor.thm.weak.DLom}; it does not make use of the scale induction.

\subsection{Resonant balls and their sparseness}
\label{ssec:res.sparse}

\bde\label{DefNR} Given $E\in\DR$ and a DSO $\BH_{\bballj(\Bx)}$,
the ball $\bballj(\Bx)$ is
called $E$-non-resonant (\ENR) if  the following bound holds:
$$
\dist\Big[ \Sigma\big(\bballj(\Bx)\big), E \Big] \ge g\delta_j .
$$
Otherwise, it will be called $E$-resonant ($E$-R).
\ede

Taking into account Theorem \ref{thm:cor.thm.weak.DLom},
we come to an important conclusion:

\bco\label{cor:no.pair.ER.L.j}
For
$g$ large enough
and any $(\om,\th)\in\Om\times\Thinf(g)$, for each given $j\ge 0$
and any $E\in\DR$, there is no pair of $3 \fN L_j$-distant \ER balls
$\bball_{L_j}(\Bx)$, $\bball_{L_j}(\By)$ $\subset$ $\bball_{L_j}^4(\Bzero)$.
\eco

\proof
Assume otherwise; then for some $3 \fN L_j$-distant balls
in $\bball_{L_j}^4(\Bzero)$
$$
\bal
& {\dist
\left[ \Sigma\left(\bball_{L_j}(\Bx);\om;\th\right),
        \Sigma\left(\bball_{L_j}(\By);\om;\th\right)
\right]}
\\
& \quad \le { \dist
\left[ \Sigma\left(\bball_{L_j}(\Bx);\om;\th\right), E\right]
+  \dist
\left[ E,\Sigma\left(\bball_{L_j}(\By);\om;\th\right)
\right]}
\\
 & \quad \le  g \delta_j  + g \delta_j  <   4 g \delta_j,
\eal
$$
which is impossible for $\th\in\Thinf(g)$, by \eqref{eq:def.Thinf}.
\qedhere

\section{Scale induction for deterministic operators}
\label{sec:MSA}

In this section, we carry out the analytic part of the multi-scale analysis of the deterministic operators
$\BH(\om;\th)$, assuming that $(\om,\th)\in \Om\times\Thinf(g)$.
The proofs are simpler and shorter than usual (in the MSA of random operators), which is understandable,
for the most tedious technical work is required to establish strong (uniform in  $\om\in\Om$)
lower bounds on spectral and inter-spectral spacings (cf. Sect.~\ref{sec:proof.key.tech.results}).

\bde

Let $L\ge \ell\ge 0$ be integers and $q\in(0,1)$. Consider a finite set $\Lam\subset\DZ^d$
such that $\Lam\supset\ball_{2L}(u)$.
A function $f:\,\Lam\to\DR_+$
is called $(\ell,q)$-dominated in $\ball_L(u)$ if for any ball
$\ball_\ell(x)\subset\ball_{2L}(u)$ one has
\be\label{eq:def.subh}
|f(x)| \leq q \;\;\mymax{y:\, |y-u| \le\ell +1} |f(y)| .
\ee
\ede

(The introduction of the ball of radius $2L$ and not $L$
is explained by the particular
strategy of proof of Lemma \ref{lem:NT.NR.is.NS} in Appendix \ref{app:lem.NT.NR.is.NS}.)

The motivation for this definition comes from the following observation.

\ble\label{lem:psi.subh}
Consider a ball $\bball=\bball_{L}(u) \subset \BLam \subseteq \bcZN$,
$L\ge \ell\ge 0$, $\Bu\in\bcZN$, and
the operator $\BH_\Lam$ with fixed potential $V$.
Fix $E\in\DR$ and let $\Bpsi\in\ell^2(\Lam)$ be a normalized eigenfunction
of $\BH_\BLam$ with eigenvalue $E$.
If every ball $\bball_{\ell}(\Bx)\subset\bball$
is \EmNS for some $m \ge 1$, then the function
$
x \mapsto |\Bpsi(\Bx)|
$
(bounded by $1$)
is $(\ell,q)$-dominated
in $\bball$, with $q = \eu^{-\gamma(m,\ell) }$.
\ele

\proof
By the GRE for the eigenfunctions (cf. \eqref{eq:GRE.EF}),
$$
|\psi(x)| \le |\ball_{\ell}(x)| \max_{y: |y-x|=\ell} |G_{\ball_\ell(x)}(x,y;E)|
\max_{y: |y-x|=\ell+1} \, |\psi(y)|.
$$
The assumed \EmNS  property of $\ball_\ell(x)\subset\ball$ implies that,
for $\ell\ge 1$, the two maxima figuring in the RHS are bounded by
$|\ball_{\ell}(x)|^{-1} \eu^{-\gamma(m,\ell)}$ and, respectively, by $\|\psi\|_\infty$.
This proves the claim.
\qedhere

\ble[Cf. Lemma 4 in \cite{C12a}]\label{lem:FG.subh}
Consider a ball $\bball=\bball_{L_{k+1}}(\Bu)$, $k\ge 0$, and the operator
$\BH_{\bball}$ with fixed potential $V$. Pick
$\Bx_0, \By_0\in\bball$ with $|\Bx_0 - \By_0| > L_{k}$, and fix $E\in\DR$.
Suppose that $\bball$ is $E$-NR and every ball $\bball_{L_{k}}(\Bx)\subset\ball$
is \EmNS for some $m \ge 1$. Then the function
$$
f_{\By_0}: \Bx \mapsto |\BG_{\bball}(\Bx, \By_0;E)|
$$
is $(L_{k},q)$-dominated
in $\bball$, with $q = \eu^{-\gamma(m,L_{k}) }$,
and bounded by $\eu^{L_k^{\beta}}$.
\ele
The proof is similar to that of Lemma \ref{lem:psi.subh} and will be omitted; the upper bound on $f_{y_0}$
follows, of course, from the \ENR property.

\ble[Cf. Lemma 2 in \cite{C12a}]\label{lem:subh.1} Let $f$ be an $(\ell,q)$-dominated function in
$\bball_L(\Bx) \subset \bball_{3L}(\Bx) \subset \BLam\subset \bcZN$.
Then
$$
 |f(\Bx)| \le q^{ \left\lfloor \frac{L+1}{\ell+1} \right\rfloor } \cM(f,\BLam)
 \le q^{ \frac{L - \ell}{\ell+1} } \cM(f,\BLam)
$$
\ele

$\blacklozenge$ We stress that the values $L=0$ and $\ell=0$ are indeed admissible.

\bde
A ball $\bballN_{j+1}(\Bu)$ is called \mbad, if for some $E\in\DR$ it contains
at least two $3 \fN L_j$-distant \EmS balls of radius $L_j$, and \mgood, otherwise.
\ede

In fact, the most important property of an $m$-good ball $\bballjone(\Bu)$
is that it contains no pair of \emph{weakly separated} balls of radius $L_j$, but
the only constructive sufficient condition for weak separability that we have
at our disposal is the lower bound ($>3\fN L_j$) on the distance between the balls.

\ble\label{lem:NT.NR.is.NS}
If a ball $\bball_{L_{j+1}}(\Bu)$, $j\ge 0$, is \mgood, with  $m\ge 1$, and
\ENR for some $E\in\DR$, then it is \EmNS.
\ele

See the proof in Appendix \ref{app:lem.NT.NR.is.NS}.

\vskip1mm

Introduce the following property which we shall prove by scale induction:

\vskip1mm
\noindent
\sparsej{}:
\textit{
For all $\th\in\Thinf(g)$, $\om\in\Om$, $E\in\DR$ and $\Bu\in\bcZN$,
there is no pair of $3 \fN L_j$-distant balls
$\bballNj(\Bx), \bballNj(\Bx)\subset \bball_{L_j^4}(\Bu)$ which are $(E,m;\om;\th)${\rm-S}.
}
\vskip2mm

Recall that we set $L_{-1} = 0$; it is convenient to formulate in a special way
the property \sparseL{L_{-1}} $\equiv$ \sparsez:

\vskip2mm
\noindent
\sparsez:
\textit{
For all $\th\in\Thinf(g)$, $\om\in\Om$, $E\in\DR$ and $\Bu\in\DZ^d$,
there is at most one  point $\Bx \in \ball_{4L_0+1}(\Bu)$ such that the single-site ball
$\bball_{0}(\Bx)$ is $(E,m;\om;\th)${\rm-S}.
}
\vskip2mm

A visible difference between \sparsez and \sparsej with $j\ge 0$ is explained by the fact
that distinct single-site "balls" are automatically weakly separated, while for the balls
of radius $L_j>0$ a minimal distance ($\ge 3\fN L_j$) is required, to guarantee their
weak separation.

\vskip1mm

For $g$ large enough (i.e.,  with $m \gg 1$), the property
\sparseL{0} follows directly from Lemma \ref{lem:sep.L0}, since
$\Thinf(g)\subset\Thzero(g)$.

\btm\label{thm:dichotomy.1} Let $\th\in\Thinf(g)$, and assume that
\sparsej holds true for some $j\ge 0$.  Then \sparsejone also holds true.
Consequently, \sparsepar{0} implies \sparsej for all $j\ge 0$.
\etm

\proof

Fix any $\th\in\Thinf$, any $\Bu\in\bcZN$ and any $E\in\DR$. Consider the ball
$\bball_{L^4_{j+1}}(\Bu)$. By definition of the set $\Th^{(j+1)}(g)\supset\Thinf(g)$
and Corollary \ref{cor:no.pair.ER.L.j},
there is no pair of weakly separated $E$-R balls of radius $L_{j+1}$ in $\ball_{L^4_{j+1}}(\Bu)$.
Let us show by contraposition that there can be no pair of weakly separated $(E,m)$-S balls
$\bball_{L_{j+1}}(\Bx)$, $\bball_{L_{j+1}}(\By)$ $\subset$ $\bball_{L^4_{j+1}}(\Bu)$.

Assume otherwise; then one of these balls -- w.l.o.g.,
let it be $\bball_{L_{j+1}}(\Bx)$ -- must
be $E$-NR. By Lemma \ref{lem:NT.NR.is.NS}, it must then contain two weakly separated $(E,m)$-S
balls of radius $L_j$, which contradicts the hypothesis \sparsej.
\qedhere

\vskip2mm
The property \sparsej, established for all length scales $L_j$, $j\ge -1$, uniformly in $\om\in\Om$,
is a stronger -- deterministic -- analog of the well-known probabilistic
"double-singularity" bound for the pairs of \EmS balls, which represents the final
result of the variable-energy MSA for random operators (cf., e.g., \cite{DK89}).

\section{Uniform localization and unimodal eigenstates}
\label{sec:Uniform}

In this section, by a Hamiltonian we always mean an operator $\BH = \BH(\om;\th)$ of the form
\eqref{eq:def.fermionic.H}.
In a number of statements, the values of the parameters $\om$ and $\th$ can be arbitrary (unless
specified otherwise), and can even be extended to a much larger class of finite-difference
operators on $\bcZN$, but some implications we prove here become meaningful only
for $\th\in\Thinf(g)$,
$g\gg 1$. As to the phase point $\om\in\Om$, recall that the key properties of the operators
$\BH(\cdot\,;\th)$ with $\th\in\Thinf(g)$ have been established uniformly in $\om$.

\subsection{Uniform decay of eigenfunctions}

\bde\label{def:loc.center}
Let $\Bpsi\in\ell^2(\bcZN)$. A point $\Bx\in\bcZN$ is called a localization center
for $\Bpsi$ iff $|\Bpsi(\Bx)| = \|\Bpsi\|_\infty$.
\ede

\bde
A (square-summable) eigenfunction $\Bpsi$ of a Hamiltonian $\BH=\BH(\om;\th)$ is called uniformly
$m$-localized if
\begin{enumerate}[\rm(a)]
  \item $\Bpsi$ has a localization center $\hBx$ such that $|\psi(\hBx)|^2 > \half \|\psi\|_2^{2}$;
  \item $\forall\, \By\in\bcZN\setminus \{\hBx\}$, one has
  $|\Bpsi(\By)| \le \eu^{-m\Brho(\Bx,\By)} \|\Bpsi\|_2$.
\end{enumerate}
When the value $m$ is irrelevant, we will say that $\Bpsi$ is uniformly localized.
\par
\noindent
A function $\Bpsi$ satisfying the condition (a) will be called unimodal.
\ede

Note that, for the operators on a countable graph $\bcZN$,
every \emph{bona fide} eigenfunction $\Bpsi$
admits a non-empty but finite set of its localization centers; it will be denoted by $\hX(\Bpsi)$.

\ble\label{lem:unimod}
\begin{enumerate}[\rm(A)]
  \item Any uniformly localized eigenfunction $\Bpsi$ of a Hamiltonian
  $\BH=\BH(\om;\th)$ has a unique localization center.

  \item Let $\{\Bpsi_j, j\in\cJ\}$, $\cJ\subset\DN$, be an orthonormal family of of uniformly localized
  eigenfunctions of a given Hamiltonian $\BH$. Then for any $\Bx\in\bcZN$, there is at most one eigenfunction $\Bpsi_i$
  with localization center $\Bx$.
\end{enumerate}

\ele

\proof
(A)
By Definition \ref{def:loc.center}, $|\Bpsi(\Bx)|$ takes the constant
value $\|\Bpsi\|_\infty$ at all its localization centers $\Bx\in\hX(\Bpsi)\ne \varnothing$.
It follows from the condition (a) of the uniform localization that
$|\Bpsi(\Bx)|^2  > \half$ for $\Bx\in\hX(\Bpsi)$.
By normalization,
$$
\bal
1 &= \sum_{\By\in\hX(\Bpsi_)} |\Bpsi(\By)|^2 + \sum_{\By\not\in\hX(\Bpsi)} |\Bpsi(\By)|^2
\ge |\hX(\Bpsi)| \cdot |\Bpsi(\hBx)|^2
> \half |\hX(\Bpsi)|,
\eal
$$
yielding $|\hX(\Bpsi)| < 2$.

\noindent
(B) Assume otherwise, and let $\Bphi, \Bpsi$ be orthogonal, normalized,
uniformly localized eigenfunctions of $\BH$ with localization center $\Bx$.
Let $\Bchi= 1 - \one_{\Bx}$.
Then we have $\|\Bchi\Bphi\|_2, \|\Bchi\Bpsi\|_2<1/2$, thus
$$
\bal
| (\Bphi,\Bpsi)| &= \Big| \Bphi(\Bx) \Bpsi(\Bx) + \sum_{\By\ne \Bx} \Bphi(\By) \Bpsi(\By) \Big|
\ge | \Bphi(\Bx)| \cdot | \Bpsi(\Bx)| - \Big| \sum_{\By\ne \Bx} \Bphi(\By) \Bpsi(\By) \Big|
\\
& > \frac{1}{\sqrt{2}} \cdot \frac{1}{\sqrt{2}} - \|\Bchi\Bphi\|_2 \, \|\Bchi\Bpsi\|_2
> \half  - \half = 0,
\eal
$$
so that $\Bphi$ and $\Bpsi$ are not orthogonal. This contradiction proves the claim.
\qedhere

\vskip1mm

In view of Lemma \ref{lem:unimod}, given an eigenbasis $\{ \psi_j\}$
of uniformly localized eigenfunctions
of a DSO $H$, we can associate with each localization center $\hBx$
of some uniformly localized eigenfunction $\psi_j$ a unique eigenvalue $\hlam = \hlam(\hBx)$ -- the one of
the eigenfunction $\Bpsi_j$.

To prove that every $\Bx\in\bcZN$ is a localization center of some eigenfunction (cf. Theorem
\ref{thm:complete.unimodal.EF}), we need the following auxiliary result.

\ble\label{lem:loc.center.S}
Let $\Bpsi$ be a normalized eigenfunction of a DSO $\BH$, and $\hBx$ any of its localization centers.
Then for any $L\in\DN$, the ball $\ball_L(\hBx)$ is $(\hlam(\hBx),m)$-S.
\ele
\proof
Fix an eigenfunction $\Bpsi$ with localization center $\hBx$ and assume otherwise.
Since $\gamma(m,L)>0$ and $q:=\eu^{-\gamma(m,L)}<1$, Lemma \ref{lem:psi.subh} implies
$$
\|\Bpsi\|_\infty =
|\Bpsi(\hBx)| \le \eu^{-\gamma(m,L)L} \max_{y\in\pt^+ \ball_L(\hBx)} |\Bpsi(y)|
   \le q \, \|\Bpsi\|_\infty
$$
 thus $\|\Bpsi\|_\infty = 0$, which is impossible, since $\|\Bpsi\|_2 >0$.
\qedhere

\btm\label{thm:EF.unif.loc}
Assume that \sparsej holds true for all $j\ge -1$, and $L_0 \ge 5$.
If, in addition, $m>0$ is large enough, so that
\be\label{eq:sum.half}
\sum_{ r\ge 1} (3  r)^{2d} \eu^{-  m  r} \le \half ,
\ee
then any normalized eigenfunction $\Bpsi$ of $\BH$ is uniformly $m$-localized.
Moreover, $\BH$ has an orthonormal eigenbasis of uniformly $m$-localized eigenfunctions.
\etm
\proof
Let us show first that any normalized eigenfunction $\Bpsi$ of $\BH$,
with localization center $\hBx$, is uniformly $m$-localized at $\hBx$.

\noindent
\textbf{Step 1.} Fix an eigenfunction $\Bpsi$ with $\|\Bpsi\|_2=1$, $\hBx\in\hX(\Bpsi)$,
$H\Bpsi = \hlam \Bpsi$,
and assume first that $R:=|\By - \hBx|\in[1,L_1]$.
By Lemma \ref{lem:loc.center.S}, the  ball $\ball_0(\hBx) = \{\hBx\}$
is $(\hlam,m)$-S.
Therefore,
by \sparseL{0},
for all $u$ with $|\hBx,u|\in[1,L_1]$, $L_1 = L_0^2$, the single-site balls $\bball_0(\Bu) = \{\Bu\}$
are $(\hlam,m)$-NS. Fix any $y$ with $1 \le |\hBx -y| \le 2L_0+1$ and set $r := R-1$.
Each single-site ball $\bball_{0}(\Bz)\subset\bball_{r}(\By)$ is
$(\hlam,m)$-NS, so by Lemma \ref{lem:subh.1}
(where we set $L=L_1$, $\ell=0$), combined with Lemma \ref{lem:psi.subh}, we have
(cf. \eqref{eq:sum.half})
$$
|\Bpsi(\By)| \le \eu^{-\gamma(m,0) \left\lfloor \frac{r+1}{0+1} \right\rfloor} \|\Bpsi\|_\infty
\le \eu^{-\gamma(m,0) (r+1)} \le \eu^{-2 m |\By - \hBx|}.
$$
Using \eqref{eq:sum.half} and the crude estimate $\card\{z:\, |z|=r\}\le (2r+1)^d \le (3r)^d$, we obtain
\be\label{eq:psi.A0}
\bal
\sum_{\By\in\bball_{2L_0}\setminus \{ \hBx \}} |\Bpsi(\By)|^2
&\le \sum_{r=1}^{2L_0} (3 r)^d \eu^{- 4 m r}.
\eal
\ee

\vskip1mm
\noindent
\textbf{Step 2.} Now consider the case where $R:=|\By-\hBx|>L_1$.
The complement of $\ball_{L_1}(\hBx)$ is covered by the disjoint annuli:
$$
\cZ \setminus \bball_{L_0}(\hBx) = \bigcup_{j \ge 2} \rA_j, \quad
\rA_j := \bball_{L_{j}}(\hBx) \setminus \bball_{L_{j-1}}(\hBx).
$$
Fix $j\ge 2$ and any $\By\in\rA_j$. The ball $\ball_{L_{j-2}}(\hBx)$ is $(\hlam,m)$-S,
and  every ball
$$
\ball_{L_{j-2}}(z)\subset \ball_{R- 3 \fN L_{j-2}}(\By)
   \subset \ball_{2 R}(\hBx) \subset \ball_{2L_{j-2}^4}(\hBx) ,
$$
is $3\fN L_{j-2}$-distant from $\ball_{L_{j-2}}(\hBx)$.
Thus $\ball_{L_{j-2}}(z)$
is weakly separated from $\bball_{L_{j-2}}(\hBx)$, by virtue of Lemma \ref{lem:3NL.WS},
and must be $(\hlam,m)$-NS. Applying again Lemma \ref{lem:psi.subh} and Lemma \ref{lem:subh.1},
with $\|\Bpsi\|_\infty \le 1$ and $C' := 3\fN + 2$, we obtain:
$$
\bal
|\Bpsi(\By)| &\le \eu^{-m(1+L_{j-2}^{-1/8}) L_{j-2} \cdot \frac{R -C'L_{j-2}}{L_{j-2}+1} } \|\Bpsi\|_\infty
\\
& \le \eu^{-m R \, \frac{1+L_{j-1}^{-1/8}}{1 + L_{j-2}^{-1}}  \cdot (1 -C'R^{-1}L_{j-2}) }
  \le \eu^{-m R \, \left(1+L_{j-1}^{-1/8} \right) \frac{ 1 - C'L^{-1}_{j-2} }{1 + L_{j-2}^{-1}}  }
\\
&
\le \eu^{-m R \, \left(1+L_{j-2}^{-1/8} \right) \left(1 - 2C'L^{-1}_{j-2} \right)  }
\le \eu^{- m R } = \eu^{- m |\By -\hBx| },
\eal
$$
provided that $L_0$ (hence, every $L_{i}$, $i\ge 0$) is large enough.
Since $|\rA_j[\le (3 L_{j})^{\fN d}$, we obtain,
with $|\By - \hBx| = R > L_j$,
\be\label{eq:psi.Aj}
\bal
\sum_{\By\in\rA_j } |\Bpsi(\By)|^2
&\le  (3 L_{j})^{2\fN d} \, \eu^{-2 m L_j}.
\eal
\ee
Collecting  \eqref{eq:psi.A0} \eqref{eq:psi.Aj} and \eqref{eq:sum.half}, we conclude that
$$
\bal
\sum_{\By\ne \hBx } |\Bpsi(\By)|^2
&\le \sum_{r=1}^{L_1} (3r)^d \eu^{-4 m r}
   + \sum_{j\ge 2} (3 L_{j})^{2\fN d} \eu^{- 2 m L_j}
\\
&\le  \eu^{-m}\, \sum_{r=1}^{\infty} (3 r)^{2\fN d} \eu^{- m r}
\le \half \,\eu^{-m} < \half.
\eal
$$
Therefore, $|\Bpsi(\hBx)|^2>1/2$, so $\Bpsi$ is uniformly $m$-localized at $\hBx$.

The above proof of uniform exponential decay of any normalized eigenfunction $\Bpsi$ is, of course,
a variant of the well-known argument, going back to \cite{FMSS85,DK89} and used
for the proof of exponential decay of any polynomially bounded solution $\BPhi$ of the equation
$\BH\BPhi = E \BPhi$. To prove the second assertion of the theorem, it suffices to repeat the Step 2,
with the following modifications:
\begin{itemize}
  \item The localization center $\hBx$ is to be replaced by any point $\hBy$ with $\BPhi(\hBy)\ne 0$.

  \item An analog of Lemma \ref{lem:loc.center.S} holds true: for $j\ge 0$ large enough,
  $\bball_{L_j}(\hBy)$ must be \EmS.

  \item The calculations of the Step 2, adapted to a polynomially bounded function $\BPhi$, prove
  its exponential decay with exponent $m>0$.
\end{itemize}
We omit the details of this well-known argument.

Finally, recall that $\BH$ is a discrete Schr\"{o}dinger operator on a graph of polynomial growth,
with bounded potential, thus, by general
results of spectral theory (cf., e.g., \cite{DisKKKR08}), for spectrally-a.e. $E\in\DR$, $\BH$
has a polynomially bounded generalized eigenfunction, and the latter, as we have shown, must
be square-summable. Moreover, by the first assertion of the theorem, such an eigenfunction
must be uniformly $m$-localized.
\qedhere

\btm\label{thm:complete.unimodal.EF}
For all sufficientlly large $m>0$ and $g\ge g_0(m)$  large enough,
so that in particular \eqref{eq:sum.half} holds true, for
any $(\th,\om)\in\Thinf(g)\times \Om$, the operator $H(\om;\th)$
has an eigenbasis of uniformly $m$-localized eigenfunctions $\Bpsi_\Bx$, uniquely
labeled by their respective localization centers:
$$
\forall\, x\in\cZ\qquad
\hX(\Bpsi_\Bx) = \{\Bx\}, \;\; |\Bpsi_\Bx(\Bx)|^2 > 1/2
$$
Consequently,
for any $\Bx\in\bcZN$ there is exactly one eigenfunction of $\BH(\om;\th)$
with localization center $\Bx$.
\etm

\proof
For all $g$ large enough,
the existence of an eigenbasis of exponentially decaying eigenfunctions $\Bpsi_k$, $k=1, 2, \ldots$,
follows from   Theorem \ref{thm:EF.unif.loc} combined with Lemma \ref{lem:sep.L0}
(proving \sparseL{L_{-1}} $\equiv$ \sparseL{0})
and Lemma \ref{thm:dichotomy.1} (inductive proof of \sparseL{L_j}, $j\ge 0$). It remains to show
that each point $\Bx$ is the localization center for exactly one eigenfunciton of $\BH(\om;\th)$.

Pick any $\Bx\in\bcZN$, then we have by the Parseval identity
$$
\bal
1 &
= \sum_k |\Bpsi_k(\Bx)|^2
= \sum_{k:\, x\in\hX(\Bpsi_k)} |\Bpsi_k(\Bx)|^2 + \sum_{k:\, \Bx\not\in\hX(\Bpsi_k)} |\Bpsi_k(\Bx)|^2
=: S_1 + S_2.
\eal
$$
For any $\th\in\Thinf(g)$, $|\hX(\Bpsi_k)|=1$, hence $\hX(\Bpsi_k)=\{\hBx(\Bpsi_k)\}$, so it remains to show
that $S_2 < 1$ in order to prove that $S_1 >0$ and, therefore, $\hBx(\Bpsi_k)=\Bx$ for exactly one
eigenfunction $\Bpsi_k$.

By Theorem \ref{thm:EF.unif.loc}, for $m>0$ large enough, any normalized eigenfunction $\Bpsi_k$
is uniformly $m$-localized, so
$$
\bal
S_2 &= \sum_{k:\, \Bx\not\in\hX(\Bpsi_k)} |\Bpsi_k(\Bx)|^2
\le \sum_{r=1}^{\infty} \;\;\sum_{k:\, |\Bx - \hBx(\Bpsi_k)| = r} \eu^{-2m r}
\le \sum_{r=1}^{\infty} (3 r)^{d} \eu^{-2m r}
\\
&
\le \frac{\eu^{-m r}}{2} < 1.
\eal
$$
Thus $1 \ge |\{k:\, \Bx\in\hX(\Bpsi_k)\}| >0$ for any $\Bx\in\bcZN$, so there exists a bijection
between the elements $\Bpsi_k$ of the eigenbasis of uniformly $m$-localized, unimodal
eigenfunctions and the vertex set $\bcZN$.
\qedhere

\subsection{Uniform dynamical localization}
\label{sec:DL.uniform}

\proof[Proof of Theorem \ref{thm:DL.uniform}]

By functional calculus,
we have the following identity, assuming that the series in the RHS of
\eqref{eq:proof.DL.1} converges absolutely:
\be\label{eq:proof.DL.1}
\langle \one_\Bx | \phi(H) | \one_\By \rangle
= \sum_{\Bz\in \DZ^d} \langle \one_\Bx  | \Bpsi_\Bz \rangle \phi(\lam_\Bz) \langle \Bpsi_\Bz| \one_\By \rangle,
\ee
so it suffices to prove convergence of the series
$$
\bal
\|\phi\|_\infty \sum_{\Bz\in \DZ^d} | \langle \one_\Bx  | \Bpsi_\Bz \rangle \langle \Bpsi_\Bz| \one_\By \rangle|
\le  \sum_{\Bz\in \DZ^d} | \langle \one_\Bx  | \Bpsi_\Bz \rangle \langle \Bpsi_\Bz| \one_\By \rangle|.
\eal
$$
By Theorem \ref{thm:complete.unimodal.EF}, we have
$
|\Bpsi_\Bz(\Bx)| \le \eu^{-m|\Bz-\Bx|}, \;\; |\Bpsi_\Bz(\By)| \le \eu^{-m|\Bz-\By|},
$
so that
$$
\bal
 \sum_{\Bz\in \DZ^d} | \langle \one_\Bx  | \Bpsi_\Bz \rangle \langle \Bpsi_\Bz| \one_\By \rangle| \le
\sum_{\Bz\in \DZ^d}  \eu^{-m |\Bx-\Bz | - m |\Bz-\By |}.
\eal
$$
Let $R = |\Bx-\By|$. For any  $\Bz\not\in\bball_{2R}(\Bx)$, we have
$$
|\Bz - \Bx| + |\Bz - \By| \ge \dist(\Bz, \bball_R(\Bx)) + \dist(\Bz, \bball_R(\Bx)) \ge 2R,
$$
since $\Bx,\By\in \bball_R(\Bx) \subset \bball_{2R}(\Bx)$. Furthermore,
$$
\forall\, n\ge R\quad \card \{\Bz\in\DZ^d:\, \dist(\Bz, \bball_R(\Bx)) = n \} \le C(d) n^{d-1},
$$
thus
$$
 \sum_{\Bz \not\in \bball_{2R}(\Bx)}  \eu^{-m |\Bx-\Bz| -m|\Bz-\By|}
 \le \sum_{n\ge 2R} C(d) n^{d-1} \eu^{-2mn} \le C'(d) R^{d} \eu^{-2mR}.
$$
For $\Bz\in\bball_{2R}(\Bx)$ (indeed, for any $\Bz\in\DZ^d$) one can use a simpler bound:
by the triangle inequality,
$
|\Bz - \Bx| + |\Bz - \By| \ge |\Bx - \By|  =R.
$
Therefore,
$$
\sum_{\Bz \in \bball_{2R}(\Bx)} \eu^{-m|\Bx-\Bz| -m|\Bz-\By|} \le  C'' R^d \eu^{-mR}.
$$
Finally,
$$
|\langle \one_\Bx | \phi(H) | \one_\By \rangle| \le \Const(d)\, |\Bx-\By|^d \eu^{-m|\Bx-\By|}.
$$
\qedhere

The standard form of dynamical localization is obtained with the functions
$\phi = \phi_t: \lam \mapsto \eu^{-\ii \lam t}$, $t\in\DR$.

\vskip5mm
\centerline{* * *}
\vskip3mm
Now the analysis of localization properties of the Hamiltonians $\BH^{(\fN)}(\om;\th)$
is completed, and we turn to the proofs of the  technical results on spectral spacings,
which made possible this analysis.

\section{Equivalence classes of the truncated interaction operators}
\label{sec:equivalence.BT}

We start with a linear-algebraic digression, postponing to Sect.~\ref{sec:proof.key.tech.results}
the analysis of the deterministic disorder.

Recall that we introduced in Sect.~\ref{ssec:BK.augmented.T} the extended kinetic
operator $\BK = \BH_0 + \BU(\Bx)$,
invariant with respect to the group of all diagonal translations
$S^a: \Bu = (u_1, \ldots, u_\fN) \mapsto (u_1 + a, \ldots, u_\fN+ a)$.
We consider the restrictions
$
\BK_{\bball_L(\Bu)} = \one_{\bball_L(\Bu)} \BK \one_{\bball_L(\Bu)}\upharpoonright\ell^2(\bball_L(\Bu)),
$
to finite balls $\bball_L(\Bu)$, and aim to prove that for each fixed $L\ge 1$,
and any $\Bu\in\bcZN$, $\BK_{\bball_L(\Bu)}$ is equivalent to on of a finite number
of operators; the notion of equivalence suitable to our purposes is defined below.

Given an integer $L\ge 1$, define the truncated interaction
$\BU_L$ generated by the truncated two-body potential
$$
U_L: r \mapsto U(r) \one_{ [0,R(L)]}(r),
$$
With $R(L)\gg 1$, $\BU_R$ is a small-norm perturbation of the full interaction $\BU$.

We specify $R(L)$ later; for the moment, fix integers $L\ge 1$ and $R=R(L)<\infty$,
and consider the truncated interaction operators $\BU_{R}$ on arbitrary $\fN$-particle
balls $\bball_L(\Bu)$, $\Bu\in\bcZN$.

Given a configuration $\Bx$ with $\Pi\Bx = \{x_1, \ldots, x_\fN\}$
(here the numeration of the pairwise distinct points of $\Pi\Bx$ is arbitrary
and introduced only as a matter of notational convenience), decompose it into the clusters
(groups) of particles with inter-cluster distance $>R$; these clusters (referred to as
$R$-clusters) will be assumed non-decomposable into smaller $R$-sub-clusters.
Let $\Gamma = \{\Gamma_1, \ldots, \Gamma_M\}$, $M_R = M_R(\Bx)$, be the entire collection
of $R$-clusters for $\Bx$. A configuration $\Bx$ with $M_R(\Bx)=1$ will be called an
$R$-monocluster configuration, or simply an $R$-monocluster.

Since the particles are indistinguishable, only the unordered collection of the cluster cardinalities,
$\fk(\Bx) = \{\fk_1, \ldots, \fk_M\} := \{|\Gamma_1|, \ldots, |\Gamma_M|\}$, is non-ambiguously defined.

\bde
Given an integer $R>0$, we say that two configurations $\Bx,\By\in\bcZN$
are $R$-equivalent iff the respective collections $\fk(\Bx)$, $\fk(\By)$ of their cluster cardinalities
are identical, as unordered collections.
\ede

It is straightforward the $R$-equivalence is indeed an equivalence relation on $\bcZN$.
The number of $R$-equivalence classes in $\bcZN$ (bounded by $\fN^\fN/\fN!$) is independent of $R$
and not important for our analysis, so we simply denote it by $\csP_\fN$ and use it
in symbolic form.

Since the inter-cluster distances are bigger than the range of the truncated interaction $\BU_R$,
the latter is completely determined by the values
$$
U(x-y) = U_R(x-y), \;\; x\in\Gamma_i \ne \Gamma_j \ni y .
$$
With this observation in mind, we introduce the following
\bde
\textbf{(i)} Two configurations $\Bx,\By\in\bcZ^{\fn}$, $\fn>1$, are called shift-equivalent
iff $\Bx = S^a \By$ for some diagonal translation $S^a$, $a\in\DZ^d$.

\textbf{(ii)} Two balls $\bball_L(\Bx),\bball_L(\By)\subset\bcZ^{\fn}$, $\fn>1$, are called shift-equivalent
iff their centers $\Bx, \By$ are shift-equivalent.

\textbf{(iii)} Two balls $\bball_L(\Bx),\bball_L(\By)\in\bcZ^{\fN}$, $\fN>1$, are called $R$-equivalent
iff $\Bx$ is $R$-equivalent to $\By$, so the two configurations have identical
$R$-cluster cardinalities $\fk_1$, \ldots, $\fk_M$, $M = M_R\ge 1$, and their clusters,
properly numbered, form the shift-equivalent pairs.
\ede

\ble
For any $L,R\ge 0$, the number of shift-equivalence classes of balls $\bball_L(\Bx)\subset\bcZ^{\fn}$
with $R$-monocluster centers $\Bx$ is bounded by $(3(\fn-1)R)^{\fn-1} \le (3\fn R)^{\fn}$.
\ele
\proof
Using the shift-invariance, we can assume without loss of generality that one of the particles
in the configuration $\Bx$ is at the origin, so it remains to assess the number of possible
$(\fn-1)$-particle subconfigurations which form, along with the particle at $0$, an $R$-monocluster.
Since the diameter of an $R$-monocluster is bounded by $(\fn-1)R$, every particle is inside the
cube of radius $(\fn-1)R$ centered at $0$; this easily gives rise to the asserted upper bound
(which, of course, is not sharp).
\qedhere

Within any cluster $\Gamma_i$, with $\fk_i = |\Gamma_i| \in[1,\fN]$, the restriction of the interaction
to set of $\fk_i$-particle configurations $\Bx_i\in\bcZ^{\fk_i}$, depends only on the relative
positions, $x-y$. This restriction, as a whole, is labeled, therefore, by $\fn_i$-tuplets
$\{x_1, \ldots, x_{\fn_i}\}$ of (distinct) points in $\DZ^d$, up to shifts of the group
as a whole:
$$
\{x_1, \ldots, x_{\fk_i}\} \mapsto \{x_1 + y, \ldots, x_{\fk_i}+y\}, \;\; y\in\DZ^d,
$$
so such a cluster is shift-equivalent to the one where, e.g., $x_1=0$. Therefore,
the number of possible, pairwise non-shift-equivalent groups is bounded
by $R^{\fk_i} \le R^\fN$. Then a rude upper bound for all non-equivalent clusterings
of $\fN$-particle configurations is by $(R^\fN)^\fN = R^{\fN^2}$, since there are at most
$\fN$ clusters.

Summarizing,

\ble\label{lem:shift.equiv.classes}
Fix any $\fN\ge 1$, $j\ge -1$, and let, as before, $L_j = L_0^{2^j}$, $R = L_j$.
There is a finite collection of $\fN$-particle configurations,
$$
\fBA_j= \{\fBa_k, k=1, \ldots, \cK_j \}, \;\; \cK_j \le \csP_\fN R^{\fN},
\;\;\csP_\fN < \infty,
$$
such that for any $\Bu\in\bcZN$, the restriction $\BK_R\upharpoonright \ell^2(\bball_L(\Bu)$
is shift-equivalent to one of operators $\BK_R\upharpoonright \ell^2(\bball_L(\fBa_k)$.
\ele

\vskip3mm
\noindent
\textbf{Example: $\fN=2$, $d=1$.} Fix $L,R\ge 0$.
There are exactly two different possible collections of cluster cardinalities:
\begin{itemize}
  \item $\fk' = \{2\}$; this is a monocluster;
  \item $\fk'' = \{1,1\}$; here we have two $1$-particle clusters.
\end{itemize}

In a monocluster, we can shift one of the particle to $0$, which leaves for the second particle
$2R$ positions: $[-R,+R]\setminus \{0\}$. Therefore, every square
$[x_1-L,x_1+L] \times [x_2-L,x_2+L]$ with overlapping projections is shift-equivalent
to one of the $2R$ squares among
$$
[-R,+R] \times [-2R, -R],  \ldots, [-R,+R] \times [R, 2R].
$$
Recall that the particles are indistinguishable, so the symmetry $(x_1,x_2)\leftrightarrow (x_2,x_1)$
leaves invariant the unordered configuration $\{x_1,x_2\}$. As a result, we only have to count the
squares
$$
Q_r := [-R,+R] \times [-R+r, +R+r],  \;\; r=0, \ldots, R.
$$
A nontrivial interaction $\BU$  on the squares $Q_r$ makes the spectrum $\Sigma(\BH_{Q_r})$ vary with
$r\in[0,R]$; at least, the spectrum might vary with $r$, and typically, it does so.
However, the $r$-dependence vanishes, once the two projections
$[-R,+R]$, $[-R+r, +R+r]$ are at distance bigger than the range of the truncated
interaction. Therefore, we encounter only a finite number of different spectral problems, while $r$
varies in the infinite lattice $\DZ^1$.

\ble
Fix any $\fN\ge 1$ and $j\ge -1$.
There exists a finite collection of configurations,
$ \fBA_j= \{\fBa_k, k=1, \ldots, \cK_j \} \subset\bcZN$,
such that for any $\Bu\in\bcZN$, $\om\in\Om$ and $\th\in\Th$, the truncated Hamiltonian
$\BH^{(\tN_j,L_j)}_{\bball_{L_j}(\Bu)}(\om;\th)$ is unitarily equivalent to one of the operators
in the finite collection
\be
\left\{ \BH^{(\tN_j,L_j)}_{\bball_{L_j}(\fBa_k)}\big(\BT^{\fBa_k}(\om, \ldots, \om);\th \big),
\;   k=1, \ldots, \cK_j \right\}
\ee
\ele

\section{Proofs of the key results on spectral spacings}
\label{sec:proof.key.tech.results}

\subsection{Entropy-type estimates in $1$-particle cubes}
\label{ssec:entropy.traj.1p}

Here we prepare ground for the combinatorial, entropy-type estimates which will be formulated
in Sect.~\ref{ssec:entropy.bounds}. Our goal is to show that in every finite ball,
there is a finite, and effectively controlled, number of scenarios which may give rise
to "small denominators", or "resonances", in the course of the scaling analysis of
the resolvents and, ultimately, eigenfunctions. The two principal mechanisms, guaranteeing
a satisfactory upper bound of unwanted events (in the parameter space $\Th$) are:
\begin{enumerate}
  \item rapid approximation of the hull $v$ by piecewise constant functions $v_N$, and
  \item tempered local divergence of trajectories $\{T^x\om\}$ (\DIV).
\end{enumerate}

Introduce some geometrical objects relative to each scale $L_j$, $j\ge 0$.
First, let
\be
R_j = \big(\, 6L_j^{4A} \,\big)^{-1}
\ee
(recall that $A,A'\in\DN^*$),
and cover the torus $\Om$ by the union of $N_{R_j}:= (R_j)^{-\nu }$ cubes $Q_{3R_j}(\om_i)$, $i\in[[1,N_{R_j}]]$, of radius $3R_j$ and with centers of the form
$$
\om_i = \left[ l_1 R_j, \ldots, l_\nu  R_j  \right),
\; l_1, \ldots, l_\nu \in [[0, (2R_j)^{-1}-1]].
$$
The order of numbering can be arbitrary. Notice that these cubes are overlapping.

Next, decompose each cube $Q_{3R_j}(\om_i)$ into a union of $3^\nu $ neighboring sub-cubes
$Q'_{R_j}(\om'_{i,k})$ of radius $R_j$, which we number starting with the central cube,
$Q'_{R_j}(\om'_{i,1})$. Observe  that the collection of all central cubes $Q'_{i,1}(R_j)$
still covers the torus $\Om$, and $\om'_{i,1} \equiv \om_i$.

Similarly, cover the torus $\Om$ by adjacent cubes $Q_{r_j}(\om''_i)$ of
radius
$$
r_j = \big(6 L_j^{4A+4A'} \big)^{-1} < R_j.
$$

\ble\label{lem:dist.r.R}
Fix $j\ge 0$ and consider $\ball_{L_j^4}(0)\subset\DZ^d$.
Fix any
point
$z\in\ball_{L^4}(0)$ and any cube $Q_{r_j}(\om_i)$. If
$T^z\om_i\in Q'_{R_j}(\om'_{\icirc,1})$ with some $i_\circ = i_\circ(i,z)$, then
\be\label{eq:lem.dist.r.R}
T^z \big(  Q_{r_j}(\om_i) \big) \subset  Q_{3R_j}(\om_{i_\circ}).
\ee
\ele

\proof
For any $\om\in Q_{r_j}(\om_i)$, we have $\dist(\om_i, \om)\le r_j$, thus by \DIV,
\be\label{eq:dist.r.R.1}
\dist( T^z \om_i, T^z\om) \le (L_j^4)^{A'} \dist(\om_i, \om) \le L_j^{4A'} r_j
\le  \frac{1}{6} L_j^{4(A' - A - A')}
=  R_j.
\ee
By assumption,
\be\label{eq:dist.r.R.2}
\dist(T^z \om_i, \om_{i_\circ}) \equiv
\dist(T^z \om_i, \om'_{i_\circ,1}) \le R_j,
\ee
therefore, by \eqref{eq:dist.r.R.1} and \eqref{eq:dist.r.R.2},
$$
\bal
\dist(T^z \om, \om_{i_\circ}) &\le
\dist(T^z \om, T^z\om'_{i}) + \dist(T^z\om'_{i}, \om'_{i_\circ}) \le R_j + R_j < 3R_j,
\eal
$$
yielding the assertion \eqref{eq:lem.dist.r.R}.
\qedhere

\bco\label{cor:finite.cover.cubes}
Let $j\ge 0$ and $n\ge 0$ be such that
\be\label{eq:6Rj.2n}
2^{-n-2} \le 6  R_j < 2^{-n-1}.
\ee
Then any cube $Q_{r_j}(\om''_\icirc)(\subset\Om)$ is covered by at most $2^\nu$
measurable sets
$\rP_{\icirc;l}$, $1\le l \le K'_{j,i_\circ} \le 2^\nu$, such that for any $z\in\ball_{L_j^4}(0)$,
$T^z \rP_{\icirc;l}$ is covered by exactly one element of the partition $\cC_n$.
\eco
\proof
Fix a cube $Q_{r_j}(\om''_\icirc)$ in $\Om$ and any $z\in\ball_{L_j^2}(0)$.
Consider the image $T^z Q_{r_j}(\om''_\icirc)$. By Lemma \ref{lem:dist.r.R}, it is covered by
one cube $Q_{3R_j}(\om_i)$, with some $i = i(\icirc)$. Since
$\diam Q_{3R_j}(\om_i)=6R_j < 2^{-n-1}$ by assumption \eqref{eq:6Rj.2n},
$Q_{3R_j}(\om_i)$ is covered by at most
$2^\nu$ adjacent cubes of side length $2^{-n}$ -- elements of the partition
$\cC_{n+1}$. As in Sect. \ref{ssec:partitions}, denote these cubes by $C_{n, k_l;i_l}$,
$l=1, \ldots, 2^\nu$. Recall that the Haar's wavelet $\ffink$ takes constant value
(either $+1$ or $-1$) on each of these cubes.

Now the claim follows by setting
$\rP_{i_\circ; l} := T^{-z} C_{n, k_l; i_l} \cap Q_{r_j}(\om''_{i_\circ})$,
$l=1, \ldots, 2^\nu$,
since each cube $Q_{r_j}(\om''_{i_\circ})$ is covered by $\le 2^\nu$
elements of the partition $\cC_{n+1}$. Naturally it suffices to retain only the non-empty intersections.
\qedhere

\vskip2mm
For each $\DZ\ni j \ge -1$, define the integers
\be\label{eq:bound.on.Nj}
\tN_j=
\begin{cases}
\tN(L_j), & j\ge 0
\\
\tN(L_0), & j = -1
\end{cases},
\qquad
\cL_j=
\begin{cases}
2^{\nu + A + A'} L_j^{4A+4A'}, & j\ge 0
\\
2^{\nu + A + A'} L_0^{4A+4A'}, & j = -1
\end{cases},
\ee
with $L  \mapsto \tN(L) = O(\ln L)$ defined in \eqref{nt.A.C}.

Further, define the operator-valued mappings
\be\label{eq:def.fh.j.positive}
\fh^{\Nj}_{j,\Bu,\th}:
\om \mapsto
\begin{cases}
\BH^{(\Nj,L_j)}_{\bball_{ L_j^4}(\Bu)}(\om;\th)\upharpoonright \ell^2(\bball_{ L_0^4}(\Bu)),
   & j\ge 0 \\
\Big(g\BV_{\tN_0}(\cdot\,;\om;\th) + \BU(\cdot) \Big) \upharpoonright \ell^2(\bball_{ L_0^4}(\Bu)), & j=-1
\end{cases} .
\ee

\ble\label{lem:loc.const.H}
Fix  $j\ge -1$ and let $\Nj = \tN(L_j)$.
Consider the $\Nj$-th approximant $v_\Nj$ of the hull $v$ given by the
expansion \eqref{eq:randelettes.1}.
For any fixed $\Bu\in\bcZN$ and $\th\in\Th$, the mapping $\fh^{\Nj}_{j,\Bu,\th}$,
defined by \eqref{eq:def.fh.j.positive}
is piecewise-constant on $\Om$.
More precisely, there is a finite collection
$\cT_j = \{\taujl, \, l=1, \ldots, \cL'_j \le \cL_j\}\subset\Om$,
with $\cL_j$ as in \eqref{eq:bound.on.Nj},
and a respective finite partition of $\Om$ into measurable subsets $\rPjl\ni \taujl$
such that $\fh^{\Nj}_{j,\th}$ is constant on each $\rPjl$.
\ele

\proof
The kinetic energy operator $\BH_0$ and the interaction operator $\BU(\cdot)$ do not depend
upon $(\om,\th)$, so we focus on the truncated potential $\BV_\tN(\om;\th)$, which is determined
by the truncated hull $v_{N_j}$.

By definition of $N_j$, the hull $v_{N_j}$ is constant on each element of the partition
$\cC_{N_j}$. By Corollary \ref{cor:finite.cover.cubes}, $\Om$ is covered by at most
$2^\nu L_j^{4A+4A'}$ measurable sets such that the image of each of them by any $T^z$,
$z\in\ball_{L_j^4}(0)$, is covered by
exactly one element of the partition $\cC_{N_j}$.
\qedhere

\subsection{Equivalence classes of the truncated interaction operators}
\vskip1mm
\noindent

Now it is convenient to introduce the operator $\BK = \BH_0 + \BU(\Bx)$; here
the disordered component $g\BV(\om;\th)$ is switched off, so $\BK$ is
invariant with respect to the group of all diagonal translations $S^a$, defined as follows:
$$
 S^a: \Bu = (u_1, \ldots, u_\fN) \mapsto (u_1 + a, \ldots, u_\fN+ a) ,
$$

We consider the restrictions of $\BK$,
$$
\BK_{\bball_L(\Bu)} = \one_{\bball_L(\Bu)} \BK \one_{\bball_L(\Bu)}\upharpoonright\ell^2(\bball_L(\Bu)),
$$
to finite balls $\bball_L(\Bu)$, and aim to prove that for each fixed $L\ge 1$,
and any $\Bu\in\bcZN$, $\BK_{\bball_L(\Bu)}$ is equivalent to on of a finite number
of operators; the notion of equivalence suitable to our purposes is defined below. Once this goal
is achieved, we will combine this result with the assertion of Lemma , thus showing that
there is a finite, reasonably large number of "scenarios" for the finite-ball operators
$\BH_{\bball_{L_j}(\Bu)}(\om;\th)$, with fixed $\th$, when $(\Bu,\om)$ runs over $\bcZN\times\Om$.

Given an integer $L\ge 1$, it will suffice for the purposes of scaling analysis to work with
the truncated interactions $\BU_L$ generated by the truncated two-body potential
$$
U_L: r \mapsto U(r) \one_{ [0,R(L)]}(r),
$$
which, for $R(L)\gg 1$, is a small-norm perturbation of the full interaction $\BU$.

We specify $R(L)$ later; for the moment, fix integers $L\ge 1$ and $R=R(L)<\infty$,
and consider the truncated interaction operators $\BU_{R}$ on arbitrary $\fN$-particle
balls $\bball_L(\Bu)$, $\Bu\in\bcZN$.

Given a configuration $\Bx$ with $\Pi\Bx = \{x_1, \ldots, x_\fN\}$
(here the numeration of the pairwise distinct points of $\Pi\Bx$ is arbitrary
and introduced only as a matter of notational convenience), decompose it into the clusters
(groups) of particles with inter-cluster distance $>R$; these clusters (referred to as
$R$-clusters) will be assumed non-decomposable into smaller $R$-sub-clusters.
Let $\Gamma = \{\Gamma_1, \ldots, \Gamma_M\}$, $M_R = M_R(\Bx)$, be the entire collection
of $R$-clusters for $\Bx$. A configuration $\Bx$ with $M_R(\Bx)=1$ will be called an
$R$-monocluster configuration, or simply an $R$-monocluster.

Since the particles are indistinguishable, only the unordered collection of the cluster cardinalities,
$\fk(\Bx) = \{\fk_1, \ldots, \fk_M\} := \{|\Gamma_1|, \ldots, |\Gamma_M|\}$, is non-ambiguously defined.

\bde
Given an integer $R>0$, we say that two configurations $\Bx,\By\in\bcZN$
are $R$-equivalent iff the respective collections $\fk(\Bx)$, $\fk(\By)$ of their cluster cardinalities
are identical, as unordered collections.
\ede

It is straightforward the $R$-equivalence is indeed an equivalence relation on $\bcZN$.
The number of $R$-equivalence classes in $\bcZN$ (bounded by $\fN^\fN/\fN!$) is independent of $R$
and not important for our analysis, so we simply denote it by $\csP_\fN$ and use it
in symbolic form.

Since the inter-cluster distances are bigger than the range of the truncated interaction $\BU_R$,
the latter is completely determined by the values
$$
U(x-y) = U_R(x-y), \;\; x\in\Gamma_i \ne \Gamma_j \ni y .
$$
With this observation in mind, we introduce the following
\bde
Two configurations $\Bx,\By\in\bcZ^{\fn}$, $\fn>1$, are called shift-equivalent
iff $\Bx = S^a \By$ for some diagonal translation $S^a$, $a\in\DZ^d$.

Two balls $\bball_L(\Bx),\bball_L(\By)\subset\bcZ^{\fn}$, $\fn>1$, are called shift-equivalent
iff their centers $\Bx, \By$ are shift-equivalent.

Two balls $\bball_L(\Bx),\bball_L(\By)\in\bcZ^{\fN}$, $\fN>1$, are called $R$-equivalent
iff $\Bx$ is $R$-equivalent to $\By$, so the two configurations have identical
$R$-cluster cardinalities $\fk_1$, \ldots, $\fk_M$, $M = M_R\ge 1$, and their clusters,
properly numbered, form the shift-equivalent pairs.
\ede

\ble
For any $L,R\ge 0$, the number of shift-equivalence classes of balls $\bball_L(\Bx)\subset\bcZ^{\fn}$
with $R$-monocluster centers $\Bx$ is bounded by $(3(\fn-1)R)^{\fn-1} \le (3\fn R)^{\fn}$.
\ele
\proof
Using the shift-invariance, we can assume without loss of generality that one of the particles
in the configuration $\Bx$ is at the origin, so it remains to assess the number of possible
$(\fn-1)$-particle subconfigurations which form, along with the particle at $0$, an $R$-monocluster.
Since the diameter of an $R$-monocluster is bounded by $(\fn-1)R$, every particle is inside the
cube of radius $(\fn-1)R$ centered at $0$; this easily gives rise to the asserted upper bound
(which is, course, not sharp).
\qedhere

Within any cluster $\Gamma_i$, with $\fk_i = |\Gamma_i| \in[1,\fN]$, the restriction of the interaction
to set of $\fk_i$-particle configurations $\Bx_i\in\bcZ^{\fk_i}$, depends only on the relative
positions, $x-y$. This restriction, as a whole, is labeled, therefore, by $\fn_i$-tuplets
$\{x_1, \ldots, x_{\fn_i}\}$ of (distinct) points in $\DZ^d$, up to shifts of the group
as a whole:
$$
\{x_1, \ldots, x_{\fk_i}\} \mapsto \{x_1 + y, \ldots, x_{\fk_i}+y\}, \;\; y\in\DZ^d,
$$
so such a cluster is shift-equivalent to the one where, e.g., $x_1=0$. Therefore,
the number of possible, pairwise non-shift-equivalent groups is bounded
by $R^{\fk_i} \le R^\fN$. Then a rude upper bound for all non-equivalent clusterings
of $\fN$-particle configurations is by $(R^\fN)^\fN = R^{\fN^2}$, since there are at most
$\fN$ clusters.

Summarizing,

\ble\label{lem:shift.equiv.classes}
... There is a finite collection of $\fN$-particle configurations,
$$
\fBA_j= \{\fBa_k, k=1, \ldots, \cK_j \}, \;\; \cK_j \le \csP_\fN R^{\fN},
$$

such that for any $\Bu\in\bcZN$,
the restriction $\BK_R\upharpoonright \ell^2(\bball_L(\Bu)$
is shift-equivalent to one of operators $\BK_R\upharpoonright \ell^2(\bball_L(\fBa_k)$.
\ele

\vskip3mm
\noindent
\textbf{Example: $\fN=2$, $d=1$.} Fix $L,R\ge 0$.
There are exactly two different possible collections of cluster cardinalities:
\begin{itemize}
  \item $\fk' = \{2\}$; this is a monocluster;
  \item $\fk'' = \{1,1\}$; here we have two $1$-particle clusters.
\end{itemize}

In a monocluster, we can shift one of the particle to $0$, which leaves for the second particle
$2R$ positions: $[-R,+R]\setminus \{0\}$. Therefore, every square
$[x_1-L,x_1+L] \times [x_2-L,x_2+L]$ with overlapping projections is shift-equivalent
to one of the $2R$ squares among
$$
[-R,+R] \times [-2R, -R],  \ldots, [-R,+R] \times [R, 2R].
$$
Recall that the particles are indistinguishable, so the symmetry $(x_1,x_2)\leftrightarrow (x_2,x_1)$
leaves invariant the unordered configuration $\{x_1,x_2\}$. As a result, we only have to count the
squares
$$
Q_r := [-R,+R] \times [-R+r, +R+r],  \;\; r=0, \ldots, R.
$$
A nontrivial interaction $\BU$  on the squares $Q_r$ makes the spectrum $\Sigma(\BH_{Q_r})$ vary with
$r\in[0,R]$; at least, the spectrum might vary with $r$, and typically, it does so.
However, the $r$-dependence vanishes, once the two projections
$[-R,+R]$, $[-R+r, +R+r]$ are at distance bigger than the range of the (truncated, thus compactly supported)
interaction. Therefore, we have only a finite number of different spectral problems, while $r$
varies in the infinite lattice $\DZ^1$.

Now consider the configurations with two $1$-particle clusters. Each single-particle cube
is an interval in $\DZ^1$ of length $2L$, and all such intervals are shift-equivalent.
Therefore, there is only one equivalence class of cubes (squares, in this case)
with $2$-cluster centers.

We conclude that there are $O(R)$ shift-equivalence classes
of $2$-particle cubes in $(\DZ^1)^2$, and the number of their images in $\bcZ^2$ is even smaller.

\vskip3mm
\noindent

Now we return to the total, but truncated, potential energy operator $\BW = g\BV + \BU$.
Combining the estimates for the components $\BV$ and $\BU$, we come to the following statement.

\ble
Fix any $\fN\ge 1$ and $j\ge -1$.
There exist
and a finite collection of configurations,
$ \fBA_j= \{\fBa_k, k=1, \ldots, \cK_j \} \subset\bcZN$,
and a finite collection of phase points,
$
(\tau^j_{l,1}, \ldots, \tau^j_{l,\fN} ) \in(\cT^j)^\fN,
$
such that
for any $(\Bu,\om)\in\bcZN\times\Om$ and any $\th\in\Th$, the truncated Hamiltonian
$\BH^{(\tN_j,L_j)}_{\bball_{L_j}(\Bu)}(\om;\th)$ is unitarily equivalent to one of the operators
in the finite collection
\be
\left\{ \BH^{(\tN_j,L_j)}_{\bball_{L_j}(\fBa_k)}(\taujl;\th), \; j=1, \ldots, \cL_j, \;
        k=1, \ldots, \cK_j \right\}
\ee
\ele

\proof Every $\Bu\in\bcZN$ is shift-equivalent to one of the points $\fBa_k$ in the collection
$\fBA_j$ constructed in Lemma \ref{lem:shift.equiv.classes}. As shows Lemma \ref{lem:shift.equiv.classes},

\qedhere

For a fixed dynamical system, the cardinality of such a collection is bounded by
$$
2^{b \tN_j^2} \cdot \big(R(L_j)\big)^{\fN^2} \le L_j^{C \ln L_j}, \;\, C = C(d,\fN).
$$

Now we fix the value $R(L) := L$. Then
$$
\|\Bu - \BU_R\| \le L^{-b' \ln L}
$$

\subsection{Entropy estimates}
\label{ssec:entropy.bounds}

Introduce the following quantities:
\begin{align}\label{eq:D.L.om.th.1a}
D^{(0)}(L, \om, \theta; \Bx, \By)
& = | \BW(\Bx;\om;\th) - \BW(\By;\om;\th) | \\
\label{eq:D.L.om.th.2a}
D^{(0)}(L, \om, \theta)
& =
\min_{ \Bx \ne \By\in\bball_{L_0^4}(\Bzero) } \;\;
D^{(0)}(L, \om, \theta; \Bx, \By), \\
\label{eq:D.L.om.th.3a}
D^{(0)}(L, \theta)
& =  \inf_{\om\in\Om} \; D^{(0)}(L, \om, \theta).
\end{align}

\proof[Proof of Theorem \ref{thm:Th.bound.Sep.H.L0}]
By definition of the quantity $\Sep{ \,\cdot\, }$,
$$
\bal
&\Sep{ \BW(\cdot\,;\om;\th), \bball_{L_0^4}(\Bzero) }
= \min_{ \Bx \ne \By\in\bball_{L_0^4}(\Bzero) }
\left| \BW(\Bx;\om;\th) - \BW(\By;\om;\th)  \right|.
\eal
$$
In contrast to Lemma \ref{lem:init.NR},
now we have to prove a bound \emph{uniform } in $\om\in\Om$. To that end, we start with
the finite subset of points $\om$ formed by the centers $\om''_{i}$ of the intervals $Q_{r_0}(\om''_i)$
(the latter cover the entire torus $\Om$).

For fixed $\taujl$ and for each pair $\Bx\ne \By$ , Lemma \ref{lem:init.NR} says that
$$
\bal
&\prth{ | \BW(\Bx;\taujl;\th) - \BW(\By;\taujl;\th) | \le 4 g \delta_0  }
\\
&\quad = \prth{ | \BW(\Bx;\taujl;\th) - \BW(\By;\taujl;\th) | \le 4 g \beta(g) 2^{-b \tN} a_\tN }
\\
   &\quad \le C L_0^{- b\tA + 8\fN d} \beta(g) .
\eal
$$
The number of pairs $(\Bx,\By)$ is bounded by
$|\bball_{L_0^4}(\Bzero)|^2\le 3^{2\fN d} L_0^{8\fN d}$,
and the number of points $\taujl$ is $\le 2 r_0^{-1} = 12 L_0^{4A+4A'}$, so the set
$$
\Th \setminus \Thzero = \{ \th:\, \Sep{ \BW(\cdot\,;\om;\th), \bball_{L_0}(\Bzero) } \le 4 g\delta_0 \}
$$
has $\prthp$-measure bounded by
$$
12 L_0^{4A+4A'} \cdot
C L_0^{8\fN d - b \tA } \beta_0(g) = C' L_0^{8\fN d + 4A + 4A' - b \tA} \beta_0(g) .
$$
\qedhere
\vskip3mm

\subsection{Separation bounds for fixed $\om\in\Om$}

\ble\label{lem:init.NR}
For all sufficiently large $g>0$ there exists
an integer $L_0 = L_0(g) \ge \eu^{c \ln^{1/2} g} >1$, a positive number
$\beta_0(g) \le \eu^{-c' \ln^{1/2}g}$, with $c, c'>0$,
such that
for any $\om\in\Om$, any $u\in\DZ^d$, with $\delta_0 = 2^{-b\tN(L_0)} a_{\tN(L_0)} \beta_0(g)$,
\be\label{eq:def.Th0.sep.V.tN}
\prth{ \Sep{ g\BV_{\tN(L_0)}(\cdot\,;\om;\th), \bball_{L_0^4}(\Bu) } < 5 g \delta_0 }
\le C L_0^{-C'} \beta_0(g)
\ee
and
\be\label{eq:def.Th0.sep.V}
\prth{ \Sep{ g\BV(\cdot\,;\om;\th), \bball_{L_0^4(\Bu)} } < 5 g \delta_0 }
\le C L_0^{-C'} \beta_0(g) .
\ee
\ele

\proof
\textbf{1. Estimates for the truncated potential.}
First of all, note that for any $\Bx\in\bball_{L_0^4}(\Bu)$ and every $x\in \Pi \Bx$,
one has $x\in \cup_{u\in \Pi\Bu} \ball_{L_0^4}(u)$. Therefore,
the function
$$
\bal
{\Bx\mapsto \BV_{\tN}(\Bx;\om;\th)\one_{\bball_{L_0^4}(\Bu)}(\Bx)}
\eal
$$
is completely determined by the sub-sample of values
$\{V_\tN(x;\om;\th), x\in \cup_{u\in \Pi\Bu} \ball_{L_0^4}(u) \}$. For this reason, we can focus
exclusively on the latter sub-sample.

Configurations $\Bu\in\bcZN$ are uniquely identified by their occupation numbers
$\Bn_z(\Bu) := \card\{ j:\, u_j = z \}$ (cf. Sect. \ref{ssec:occup.numbers}).
Fix any $\Bx \ne \By$ and
$\om\in\Om$.
We have
$$
\bal
\BV_\tN(\Bx;\om;\th) - \BV_\tN(\By;\om;\th) &= \sum_{u\in\Pi \Bx \cup \Pi \By}
\left(\Bn_u(\Bx)  - \Bn_u(\By)  \right) \, V_\tN(u;\om;\th)
\\
& = \sum_{u\in\Pi \Bx \cup \Pi \By} \Bc_{\Bx,\By}(u) \, V_\tN(u;\om;\th),
\eal
$$
where at least one coefficient
\be\label{eq:c.x.y}
\Bc_{\Bx,\By}(u) := \Bn_u(\Bx)  - \Bn_u(\By)\in\{-1, 0, 1\}
\ee
is nonzero,
since otherwise we would have $\Bx=\By$.

For any fixed $u$ and $\om$ and any $l\ge 0$,
there is a unique integer $\hk_l(T^u\om)\in [1,K_n]$ such that (cf. \eqref{eq:hk})
$$
v_\tN(T^u\om;\th) = a_\tN \th_{\tN, \hk_\tN(T^u\om)} +
\sum_{l < \tN} \sum_{k=1}^{K_{l}} a_{l} \th_{l,\hk_l(T^u\om)}
=: a_\tN \th_{\tN, \hk(u)} + \hth_{\tN,u}(\om;\th),
$$
where $\hth_{\tN,u}(\om;\th)$ and $\th_{\tN, \hk_\tN(T^u\om)}$ are independent for fixed $\om$,
as random variables on $(\Th,\fB,\prthp)$.
Moreover, for any $u, u'\in\DZ^d$, the random variables
$\hth_{\tN,u'}(\om;\th)$ and $\th_{\tN, \hk_\tN(T^u\om)}$ are also independent (for fixed $\om$).

By definition of $\tNL$ (cf. \eqref{nt.A.C} and the assumption \LVB,
the elements of the partition $\cC_\tN$ separate all points
$\{T^u \om,\; u\in\ball_{L_0^4}(0)\}$,
Since $\Bx,\By\in\bball_{L_0^4}(\Bzero)$,  all points of the set
$
\{ T^u\om,\; u\in \Pi \Bx \cup \Pi \By\}
$
are separated by the elements of $\cC_{\tN(L_0^4)}$.
Therefore, with $\Bc_{\Bx,\By}$ defined in \eqref{eq:c.x.y}, we have
$$
\bal
\BV_\tN(\Bx;\om;\th) - \BV_\tN(\By;\om;\th) &=
\BX(\om;\th) := \sum_{u\in\Pi \Bx \cup \Pi \By}
\Bc_{\Bx,\By}(u) \,  v(T^u\om;\th)
\\
& =\sum_{u\in\Pi \Bx \cup \Pi \By}
\Bc_{\Bx,\By}(u) \, \left( a_\tN \th_{\tN,k(u)} + \zeta_{\tN}(T^u\om;\th) \right)
\\
& = a_\tN\BY(\om;\th) + \BZ(\om;\th),
\eal
$$
where
$$
\bal
\BY(\om;\th) &:=  \sum_{u\in\Pi \Bx \cup \Pi \By}
\Bc_{\Bx,\By}(u) \,  \th_{\tN,k(u)},
\\
\BZ_{\tN, \Bx,\By}(\om;\th) &:= a_\tN \sum_{u\in\Pi \Bx \cup \Pi \By}
\Bc_{\Bx,\By}(u) \,  \zeta_{\tN}(T^u\om;\th).
\eal
$$
Observe that $\BZ(\om;\th)$ is independent of $\BY(\om;\th)$.

Now consider the sigma-algebra $\fB_{L^4_0}$, figuring in
the property $\LVBU$ (we take $\fB_L$ with $L = L^4_0$).
Conditional on $\fB_{L_0^4}$, the random variables
$\th_{\tN,k(u)}\sim \Unif([0,1])$
are IID,
so their sum $\BY(\om;\th)$, with coefficients $\Bc_{\Bx,\Bu}(u)\in\{0,\pm1\}$,
among which at least one is nonzero, has probability density bounded by $1$.
On the other hand, $\BZ(\om;\th)$ is $\fB_0$-measurable, so we obtain
$$
\bal
\prth{ |g\BX(\om;\th)| \le gs\, \big|\, \fB_{0} }
&=\prth{ |a_{\tN} \BY(\om;\th) + \BZ(\om;\th)| \le s \,\big|\, \fB_{0} }
\\
& \le \sup_{t\in\DR} \, \prth{ \big| a_{\tN}\BY(\om;\th) - t\big| \le s \,\big|\, \fB_{0} }
\\
&  \le 2 a_{\tN}^{-1} s
\eal
$$
(the above equalities and inequalities hold $\prthp$-a.s.).
Thus
$$
\bal
\prth{ |g\BX(\om;\th)| \le gs }
= \esmth{ \prth{ |g\BX(\om;\th)| \le gs \,\big|\, \fB_{0}} }
%
\le 2 a_{\tN}^{-1} s .
\eal
$$
Counting the number of pairs $\Bx, \By\in \bball_{L_0^4}$, we conclude that
$$
\prth{ \min_{\Bx\ne\By \in \bball_{L_0^4}} \big| g\BV(\Bx;\om;\th) - g\BV(\By;\om;\th) \big| < gs }
\le C L_0^{ 8\fN\, d } a_{\tN}^{-1} s .
$$

Let $\beta>0$ (the appropriate value of $\beta$ will be chosen later) and
$$
s = 5 \beta  a_{\tN}.
$$
Then, with $\tA$ defined in \eqref{eq:def.tA},
\be\label{eq:prth.Th.0.om}
\bal
\prth{ \Sep{ gV_\tN(\cdot;\om;\cdot), \bball_{L_0^4}}  < 5g \beta a_\tN }
\le C L_0^{8\fN d} \beta \,  \le C \beta \, L_0^{8\fN d} L_0^{-4 b\tA },
\eal
\ee
Let
\be
\Thzero(g,\om) :=
\left\{\th\in\Th:\,  \Sep{ gV_\tN(\cdot;\om;\th), \bball_{L_0^4}} \ge 5g \beta 2^{-b\tN } a_\tN \right\},
\ee
then by \eqref{eq:prth.Th.0.om}, we have
\be\label{eq:prthp.Thzero}
\bal
\forall\, \th\in\Thzero(g,\om) \;\;
\Sep{ g\BV_\tN(\cdot;\om;\th), \bball_{L_0^4}}  &\ge 5 g \beta a_\tN \\
\eal
\ee
and
$$
\prth{\Thzero(g,\om)}  \ge 1 - C  \, 2^{8\fN d } \beta.
$$
On the other hand,
$$
\| gV - gV_\tN\|_\infty \le g 2^{- 2b\tN} a_\tN,
$$
Set
$$
\beta = \beta(L_0) := 2\fN \cdot 2^{-2 b \tN(L_0)}
$$
then
$$
\| g\BV - g\BV_\tN\|_\infty \le \fN\cdot g 2^{- 2b\tN} a_\tN = \half g\delta_0.
$$
Thus for any $\th\in\Thzero(g,\om)$, by the triangle inequality,
$$
\bal
\Sep{ g\BV(\cdot;\om;\th), \bball_{L_0^4}}
  &\ge \Sep{ g\BV_\tN, \bball_{L_0^4}} - 2 \| g\BV - g\BV_\tN\|_\infty
\\
& \ge 5 g \delta_0 - 2 \cdot \half g\delta_0 = 4 g \delta_0.
\eal
$$

Given any $m\ge 1$ and $L_0\ge 1$, assume that
$$
g \ge g_0(m, L_0) := 2 \fN d \,\eu^{4m} \, 2^{b\tN } a^{-1}_{\tN}
\equiv  \eu^{4m} \, 2^{b \tN^2(L^4_0) + 2b\tN(L^4_0)  } ,
$$
so that
\be\label{eq:Sep.5.e.m}
\Sep{ g\BV, \bball_{L_0^4}} \ge 16 \fN d\, \eu^{4m}, \; m \ge 1.
\ee
Clearly, as $g\to+\infty$, the maximum value of $L_0$ we can afford in the above
inequalities also tends to $+\infty$. Set
\be\label{eq:choice.L_0.g}
\bal
L_0(g) &:=
\max \left\{L\in \DN^*:\;g \ge \eu^{4m} \, 2^{b \tN^2(L^4) + 2b\tN(L^4) } \right\},
\\
\beta_0(g) &:= \beta(L_0(g)).
\eal
\ee
Then it is readily seen that
\be\label{eq:beta.L0}
\lim_{g\to\infty} L_0(g) = +\infty, \;\;
\lim_{g\to+\infty} \beta_0(g) = 0.
\ee
More precisely, by definition of $L \mapsto \tN(L)$, we have, with some $\tA, c, c', c''>0$,
$$
\bal
\tA \ln L_0 = \tN(L_0) & \sim \frac{\ln^{1/2} g}{ (2b)^{1/2}} ,
\\
\ln L_0(g) & \sim c \ln^{1/2} g ,
\\
\beta_0(g) &\sim L_0^{-c' \tA } \sim \eu^{ -c'' \ln^{1/2} g } .
\eal
$$

Now, with $\delta_0 = \beta_0(g) 2^{-b \tN} a_\tN$ as in \eqref{eq:def.delta.j}, and $\tA$ as in
\eqref{eq:def.tA},
\be\label{eq:final.sep.bound.L0}
\bal
\prth{ \Thzero(g, \om)} \equiv \prth{ \Sep{g\BV_\tN, \bball_{L_0^4} } \ge 5 g \delta_0}
&\ge 1 - C L_0^{8 \fN d} \beta_0(g)
\\
&\ge  1 - C L_0^{- 8 b A + 8 \fN d} .
\eal
\ee
We have shown that the inequality $\Sep{g\BV_\tN, \bball_{L_0^4} } \ge 5 g \delta_0$ for the truncated
potential implies a slightly weaker bound
$\Sep{g\BV, \bball_{L_0^4} } \ge 4 g \delta_0$ for the original one,
thus the asserted bounds \eqref{eq:def.Th0.sep.V.tN}--\eqref{eq:def.Th0.sep.V}
follow from \eqref{eq:final.sep.bound.L0}.
\qedhere

\vskip1mm
\noindent
\subsection{Spectral spacings for single-site balls (scale $L_{-1}$).}

\bde
Given a ball $\bball_L(\Bx)$ and a sample operator $\BH_{\bball_L(\Bx)}(\om;\th)$,
we say that $\bball_L(\Bx)$ is $(E,m)$-non-singular (\EmNS), for $E\in\DR$, $m>0$,
if for any $\By\in\pt^- \bball_L(\Bx)$
the resolvent $\BG_{\bball_L(\Bx)}(E) = \left( \BH_{\bball_L(\Bx)} - E\right)^{-1}$
admits the upper bound
\be\label{eq:def.EmNS}
|\BG_{\bball_L(\Bx)}(\Bx,\By;E)| \le
\left\{
  \begin{array}{ll}
   (3L)^{- \fN d}\, \eu^{-\gamma(m,L)}, & \hbox{if $L\ge 1,$} \\
   (2\fN d)^{-1}\eu^{-\gamma(m,L)}, & \hbox{if $L=0,$}
  \end{array}
\right.
\ee
(here it is assumed implicitly that $E \not\in \Sigma(\bball_L(\Bx))$), with
\be\label{eq:def.gamma}
\gamma(m,L) :=
\left\{
  \begin{array}{ll}
   m(1 + L^{-1/8})L , & \hbox{if $L\ge 1$,} \\
   2m, & \hbox{if $L=0$.}
  \end{array}
\right.
\ee
Otherwise, $\bball_L(\Bx)$ is called $(E,m)$-singular (\EmS).
\ede

Notice that for any $L\ge 1$ and any $m>0$,
$mL < \gamma(m,L)  <2mL$.

\ble\label{lem:sep.L0}
Let $\th\in\Thzero(g)$, $g \ge (\eu^{2mL_0}+2d)^2$, $m>0$.
Then for any $\om\in\Om$, any $\Bu\in\bcZN$ and any $E\in\DR$, there is at most one
single-site ball $\{\Bx\} = \bball_{0}(\Bx)\subset \bball_{L_0^4}(\Bu)$ which is \EmS.
\ele
\proof
It follows from the definition of the subset
$\Th^{(0)}(g)$ (cf. \eqref{eq:def.Th0.sep.V}) that,
for $\Bx,\By\in\ball_{L^4_0}(\Bzero)$ with $\Bx\ne \By$,
$$
| \BW(\Bx;\om;\th) - \BW(\By;\om;\th) + 2d)| \ge 4 g \delta_0 > 16 \fN d\eu^{4 m},
$$
thus for any $E\in\DR$, for at least one $\Bz\in\{\Bx, \By\}$, we have
$$
\bal
\| \BG_{\bball_0(\Bz)}(E)\| &= \big| \left(\BW(\Bz;\om;\th) - E \right)^{-1}   \big| \le
\le (8\fN d)^{-1} \eu^{-4m}
\\
& < (2\fN d)^{-1} \eu^{-\gamma(m,0)} ,
\eal
$$
yielding the \EmNS property of $\bball_{0}(\Bz)$. Therefore, no pair of distinct single-site balls
$\bball_0(\Bx)$, $\bball_0(\By)\subset\bball_{L_0^4}(\Bu)$ can be \EmS.
\qedhere

\subsection{"Entropy" estimates for $\BH(\cdot\,;\th)$}
\label{ssec:entro}

\bco\label{cor:loc.const.H}
Fix any $j\ge 0$ and let $\tN_j = \tN(L_j)$.
Consider the $\tN_j$-th approximant $v_{\tN_j}$ of the hull $v$ given by  \eqref{eq:randelettes.1}.
For any fixed $\th\in\Th$, the mapping
$\fh^{\tN_j}_{j,\th}$ defined in \eqref{eq:def.fh.j.positive}
is piecewise-constant on $\Om$.
More precisely, $\Om$ is covered by at most $\cL_j=2^\nu L_j^{4A+4A'}$ measurable sets
$\rPjl$
(uniquely labeled by some points $\taujl\in\rPjl$)
on which $\fh^{\tN_j}_{j,\th}$ is constant. Therefore,
the operator-valued function
$\fh^{\tN_j}_{j,\th}$
takes on $\Om$
only a finite number of values, among
\be
\BH^{(\tN_j)}_{\bball_{L_j^4}(\Bzero)}(\taujl;\th), \;\; 1 \le l \le l_{\rm max}(j) \le \cL_j.
\ee
\eco
\proof
By definition of $L\mapsto \tNL$ (cf. \eqref{nt.A.C}), the hull $v_{\tN_j}$
is constant on each element of the partition $\cC_{\tN_j}$.
By Corollary \ref{cor:finite.cover.cubes}, $\Om$ is covered by at most
$2^\nu L_j^{4A+4A'}$ measurable sets such that the image of each of them by any $T^z$,
$z\in\bball_{L_j^4}(\Bzero)$, is covered by
exactly one element of the partition $\cC_{\tN_j}$.
\qedhere

\vskip2mm

Introduce the quantities, analogous to those defined in \eqref{eq:D.L.om.th.1a}--\eqref{eq:D.L.om.th.3a}:
\begin{align}\label{eq:D.N.L.om.th.1}
D^{(N)}(L, \om, \theta; \Bx, \By)
& = \dist\left(\Sigma\big(\bball_L(\Bx); \om;\th\big),
    \; \Sigma\big(\bball_L(\By);\om;\th \big) \right),
\\
\label{eq:D.N.L.om.th.2}
D^{(N)}(L, \om, \theta)
& =  \min_{ \lr{\bball_L(\Bx),\bball_L(\By)} \sqsubset \bball_{L^{4}}(\Bzero) }
\;\;
D^{(N)}(L, \om, \theta; x, y),
\\
\label{eq:D.N.L.om.th.3}
D^{(N)}(L, \theta)
& =  \inf_{\om\in\Om} \; D^{(N)}(L, \om, \theta).
\end{align}

\bco\label{cor:approx.H.const}
Fix $j\ge 0$. Using the notations of Corollary \ref{cor:loc.const.H} and
\eqref{eq:D.N.L.om.th.1}--\eqref{eq:D.N.L.om.th.3},
with $\tN_j = \tN(L_j)$,
assume that
for each $\taujl$ one has (cf. \eqref{eq:def.delta.j})
$$
D^{(\tN_j)}(L_j, \taujl, \th) \ge 5 g \delta_j .
$$
Then for the original operator $\BH_{\bball_{L_j^4}(\Bzero)}(\om;\th)$, one has the uniform
lower
bound
$$
D(L_j,\th) = \inf_{\om\in\Om} \; D(L_j, \om, \th) \ge 4g \delta_j.
$$
\eco
\proof
The claim follows from Corollary \ref{cor:loc.const.H} combined with the norm-bound
for the perturbation $\BH - \BH^{(N)}$, following from \eqref{eq:aN.decay.fast}:
for any finite subset $\BLam\subset\bcZN$,
$$
\| \BH_\BLam(\om;\th) - \BH_\BLam^{(\tN_j)}(\om;\th)\| \le g\delta_j,
$$
and the min-max principle for the eigenvalues of self-adjoint operators, since
the perturbation of
each eigenvalue of $\BH_\BLam(\om;\th)$, due to the approximation by $\BH_\BLam^{(\tN_j)}(\om;\th)$,
is bounded by $\| \BH_\BLam(\om;\th) - \BH_\BLam^{(\tN_j)}(\om;\th)\|$.
\qedhere

We see that in order to guarantee a lower bound on $D(L_j,\th)$, it suffices
to estimate a finite number of $\th$-probabilities for the approximants of order
$\tN_j$ and $\om \in\{ \taujl, 1 \le l\le \cL_j\}$. This task is performed in the next subsection.

\subsection{Exclusion of bad $\th$-sets by the Wegner estimate}
\label{ssec:proof.thm:cor.thm.weak.DLom}

\proof[Proof of Theorem \ref{thm:cor.thm.weak.DLom}]
Fix $j\ge 0$ and let $\tN_j=\tN(L_j) = O(\ln L_j)$ be given by \eqref{nt.A.C}.
Further, fix a pair of
disjoint balls $\bball_{L_j}(x)$, $\bball_{L_j}(y) \subset \bball_{L_j^4}(\Bzero)$ and consider
the operators $\BH^{(N)}_{\bball_{L_j}(x)}$, $\BH^{(\tN_j)}_{\bball_{L_j}(y)}$.
Given a ball $\bball_{L_j}(\Bu)$, $\Bu\in\bcZN$, we denote by $\SigmaNj(\bball_{L_j}(\Bu);\om;\th)$
the spectrum of $\BH^{(\tN_j)}_{\bball_{L_j}(\Bu)}(\om;\th)$.
\vskip1mm

Recall that all the points of any finite trajectory of the form
$\{T^z\om, \, z\in\bball_{L_j^4}(\Bzero)\}$ are separated by the elements $C_{\tN_j,k}$
of the partition $\cC_{\tN_j}$. Such a separation occurs in particular for
$\{T^z\om, z\in \ball_{L_j}(x) \cup \ball_{L_j}(y)\}$,
thus conditional on the sigma-algebra $\cB_{\ne \tN_j}$ generated by
$\{ \th_{n,k}, \, n\ne \tN_j\}$, and with fixed $\om\in\Om$,
the probability distribution of the potential $V_{\tN_j}(z;\om,\th)$, generated by the truncated hull
$v_{\tN_j}$, gives rise to the sample
of independent random variables (relative to the probability space $\Th$, and not $\Om$ !)
\be\label{eq:cV}
\cV_{\tN_j}(\om,\th) := \{ v_{\tN_j}(T^z;\om,\th), z\in \ball_{L^4}(\Bzero) \};
\ee
each of them is uniformly distributed  in its individual interval
$[c_z, c_z +a_{\tN_j}]$, with $c_z = c_z(\om,\th)$ determined by the
random (in $\th$) amplitudes $\th_{n,k}$, from generations with $n < \tN_j$.

Therefore, conditional on $\cB_{\ne \tN_j}$, the independent random variables listed in \eqref{eq:cV}
have individual probability densities, uniformly bounded by $a_{\tN_j}^{-1}$. As a result,
Theorem \ref{thm:weak.sep.Wegner}
applies to the operators
$\BH^{(N_j)}_{\bball_{L_j}(\Bx)}$ and $\BH^{(\tN_j)}_{\bball_{L_j}(y)}$.
Specifically,
the assertion \eqref{eq:Thm.Wegner.WS}
of Theorem \ref{thm:weak.sep.Wegner}
implies that, for every fixed $\taujl$
$$
\bal
&\prTh{ D^{(\tN_j)}(L, \taujl, \theta; \Bx, \By) \le g \cdot 5 \delta_j  }
\\
& \quad
\equiv \prTh{ \dist
\left[ \SigmaNj(\ball_{L_j}(\Bx); \taujl;\th),\;
        \SigmaNj(\ball_{L_j}(\By); \taujl;\th)
\right] \le g\cdot 5\delta_j  }
\\
& \quad \le C_5 L_j^{2B \ln L_j} \, (5 \delta_j)^{2/3}
\le C_6 L_j^{2B \ln L_j} \, \delta_j^{2/3} .
\eal
$$
The number of all pairs
$\Bx,\By\in\ball_{L_j^4}(\Bzero)$ is bounded
by $|\bball_{L_j^4}(\Bzero)|^2/2 \le 3^{2\fN d} L_j^{4\fN d}/2$, so
we obtain (cf. \eqref{eq:D.N.L.om.th.2})
$$
\prTh{ D^{(\tN_j)}(L, \taujl, \theta) \le 5 \delta_j }
\le \half 3^{4\fN d} L_j^{8\fN d} C_6 L_j^{2B \ln L_j} \, \delta_j^{2/3}.
$$
Further,
$$
\bal
\prTh{ \min_l D^{(\tN_j)}(L, \taujl, \theta) < 5g \delta_j}
&\le \cL_j \min_l \prTh{  D^{(\tN_j)}(L, \taujl, \theta) < 5 g \delta_j}
\\
& \le C(d,\nu) C_6 L_j^{8\fN d +4A' + 2B \ln L_j} \, \delta_j^{2/3} .
\eal
$$
By Corollary \ref{cor:approx.H.const}, we conclude that
$$
\bal
\prTh{ \inf_{\om\in\Om} D(L_j,\th) < 4g \delta_j} & \le
\prTh{ \min_{l} D^{(\tN_j)}(L, \taujl, \theta) < 5 g \delta_j}
\\
&\le C_1(d,\nu) L_j^{8d + 4 A + 4 A'} \,\delta_j^{2/3}.
\eal
$$
By construction (cf. \eqref{eq:def.delta.j}),
$\delta_j = 2^{-2b \tN_j} a_{\tN_j} = C''  L_j^{-2 b A} a_{N_j}$,
so
$$
\bal
\prTh{ \inf_{\om\in\Om} D(L_j,\th) < 4 g \delta_j} & \le
 C_2(\fN, d,\nu) g^{-1/2} L_j^{8\fN d + 4 A + 4 A' - \frac{4}{3}b A }
\\
&\le  C_2(\fN, d,\nu)  L_j^{- b A },
\eal
$$
and since we have assumed $b > (24\fN d + 12A + 12A')/A$,
for $L_0$ (hence, all $L_j$) large enough (or for even larger $b>0$),
we get a simpler upper bound: for all $j\ge 0$,
\be\label{eq:mu.Th.j.proof}
\bal
\prTh{ \inf_{\om\in\Om} D(L_j,\th) < 4 g \delta_j} & \le    L_j^{- bA}.
\eal
\ee
\qedhere

Now define the sets
\be\label{eq:def.Thinf.again}
\bal
\Th^{(j)}(g) &:= \myset{\th\in\Th:\,  \inf_{\om\in\Om} D(L_j,\th) \ge 4g \delta_j },
\\
\Thinf(g) &:= \cap_{j\ge 0} \Th^{(j)}(g).
\eal
\ee

\appendix

\section{ Geometric resolvent equations}
\label{sec:App.GRE}

Recall some standard facts about the DSO on locally finite graphs,
$H_{\Lam'} = -\Delta_{\Lam'} + V$, where $\Lam'\subset\cG$
is a finite connected lattice subset partitioned into $\Lam \sqcup \Lam^\rc$,
$\Lam^\rc := \Lam' \setminus \Lam^\rc$.
The structure of the graph is irrelevant for the general results considered in this subsection,
applicable, in particular, to the fermionic Hamiltonians $\BH$ on subgraphs of $\bcZN$. For this reason,
we abandon in this subsection our usual boldface notation.

The graph Laplacian in $\Lam'$, considered as
the canonical graph Laplacian (the edges of $\Lam'$ being formed by the nearest-neighbor
pairs $(x,y)$) admits the decomposition
$
\Delta_{\Lam'} = \Delta_{\Lam} \oplus \Delta_{\Lam^\rc} + \Gamma_{\Lam, \Lam^\rc}
$,
where $\Gamma_{\Lam, \Lam^\rc}$ has the form
$$
\Gamma_{\Lam, \Lam^\rc} =
\sum_{(z,z')\in\pt\Lam} \big(\, \ket{\one_z} \bra{\one_{z'}} + \ket{\one_{z'}} \bra{\one_{z}} \,\big)
$$
(we use the standard Dirac's "bra-ket" notations). Similarly,
$$
H_{\Lam'} = H^\bullet_{\Lam, \Lam^\rc} - \Gamma_{\Lam, \Lam^\rc}, \;\;
H^\bullet_{\Lam, \Lam^\rc}= H_{\Lam} \oplus H_{\Lam^\rc}.
$$
Let $G^\bullet_{\Lam, \Lam^\rc}(E) = (H^\bullet_{\Lam, \Lam^\rc} - E)^{-1}$,
$G_{\Lam'}(E) = (H_{\Lam'} - E)^{-1}$,
$G_{\Lam}(E) = (H_{\Lam} - E)^{-1}$.
The second resolvent equation implies (cf., e.g., \cite{CL90,DisKKKR08})
$$
G_{\Lam'}(E) = G^\bullet_{\Lam, \Lam^\rc}(E)  +
G^\bullet_{\Lam, \Lam^\rc}(E) \Gamma_{\Lam, \Lam^\rc} G_{\Lam'}(E).
$$
For $x\in\Lam, y\in\Lam^\rc$, we obtain the \emph{geometric resolvent equation} (GRE):
\be\label{eq:GRE.GF}
G_{\Lam'}(x,y;E) = \sum_{(z,z')\in\pt \Lam}
G_{\Lam}(x,z;E) G_{\Lam'}(z',y;E).
\ee
Similarly, for a solution to the equation $H_{\Lam'}\psi = E \psi$ with
$E\not\in \Sigma( H_{\Lam})$, one obtains (cf. e.g., \cite{CL90,DisKKKR08}), for any $x\in \Lam$,
the GRE for eigenfunctions
\be\label{eq:GRE.EF}
\psi(x) = \sum_{(z,z')\in\pt \Lam} G_{\Lam}(x,z;E) \psi(z').
\ee

\section{Proof of Lemma \ref{lem:3NL.WS}}
\label{sec:proof.Lemma.lem:3NL.WS}

Call a configuration $\Bx\in\bcZN$ $R$-decoupled if it can be decomposed into a union
of two complementary sub-configurations, $\Bx'\in\bcZ^{(\fn')}$ and $\Bx''\in\bcZ^{(\fn'')}$,
$\fn'+\fn''=\fN$, such that $\rd_\cZ(\Pi \Bx', \Pi \Bx'') > R$.

Next, call an $R$-cluster of the configuration $\Bx\in\bcZN$ any sub-configuration
$\Bx_\cJ$ which is not $R$-decoupled and is not contained in any strictly larger non-$R$-decoupled
sub-configuration $\Bx'$.

It follows  from the triangle inequality that the diameter of any $R$-cluster of $\Bx\in\bcZN$
is upper-bounded by $2(\fN-1) R$. Clearly, any configuration $\Bx$ can be decomposed into a
disjoint union of $R$-clusters, with distance $>R$ between the clusters, if $\Bx$ is not itself
a single $R$-cluster.

Let $\BGamma(\Bx,2L)= \{\Gamma_1, \ldots, \Gamma_M \}$
be the collection of $2L$-clusters of  $\Bx$.

Although the particles are considered to be indistinguishable and there is no canonical order
on the points of $\Pi \Bx$, it is convenient now to numerate them in some way, so
$\Gamma_i = \{x_j, j\in \cJ_i\}$, where $\cJ_1, \ldots, \cJ_M$ form a partition of $[[1,\fN]]$.
Numerate also in some way the particles of the configuration $\By$.

Further, consider the $L$-neighborhood of each cluster $\Gamma_i$, i.e. the union
$$
\cU_i := \cup_{j\in\cJ_i} \ball_L(x_j)
$$
and find some minimal balls $Q_1,\ldots, Q_M$ containing the respective unions
$\cU_1, \ldots, \cU_M$. By minimal balls  we mean balls of the minimal possible radius;
$Q_i$ is not necessarily uniquely defined. It is straightforward that
$$
\diam( Q_i ) \le \diam( \Gamma_i ) + 2L \le 2(\fN - 1)L + 2L = 2 \fN L.
$$
Introduce the occupation numbers of the balls $Q_i$ for the configurations $\Bx$
and $\By$:
$$
\bal
n_i(\Bx) &:= \card( \Pi \Bx \cap Q_i), \;\; i=1, \ldots, M,
\\
n_i(\By) &:= \card( \Pi \By \cap Q_i), \;\; i=1, \ldots, M.
\eal
$$
There are two possibilities:

\noindent
(I) For all $1\le i \le M$, we have $n_i(\Bx) = n_i(\By)$. Then there exists a permutation
$\pi \in\fS_\fN$ such that for all $1\le j \le \fN$,
$$
\rd_\cZ(x_{\pi(j)}, y_j ) \le \diam( Q_i) \le 2\fN L.
$$
This permutation is a mere artefact of the numeration which we introduced arbitrarily in
$\Pi \Bx$ and $\Pi \By$. In terms of the indistinguishable particle configurations, we simply have
$$
\Brho(\Bx, \By) \le 2 \fN L,
$$
so this situation (identical occupation numbers,  $n_i(\Bx) = n_i(\By)$, $1 \le i \le M$)
can be ruled out under the assumption $\Brho(\Bx, \By) > 2 \fN L$.

\noindent
(II) For some $1\le i \le M$,  $n_i(\Bx) \ne n_i(\By)$. By construction of $\Gamma_i$, it contains
$|\Gamma_i|\ge 1$ particles from $\Bx$, so $n_i(\Bx)\ge 1$ for all $i$. Observe that
$$
\sum_{i=1}^M \big( n_i(\Bx) - n_i(\By) \big) = \fN - \sum_{i=1}^M n_i(\By) \ge 0,
$$
since each of the $\fN$ particles from $\Bx$ belongs to exactly one cluster; this is not
necessarily true for the particles from $\By$ (all or any number of which can be outside all balls $Q_i$),
but we still have $\sum_i n_i(\By) \le \fN$.

Since not all quantities $\big( n_i(\Bx) - n_i(\By) \big)$ vanish, there exists some
$i_\circ\in[[1,M]]$ such that $\big( n_{i_\circ}(\Bx) - n_{i_\circ}(\By) \big)>0$: otherwise,
the above LHS would be negative.

Setting now $Q = Q_{i_\circ}$, we see that the conditions \eqref{eq:def.weak.sep} are fulfilled, i.e., $\bball_L(\Bx)$
and $\bball_L(\By)$ are weakly $Q$-separated.
\qedhere

\section{ Proof of Theorem \ref{thm:weak.sep.Wegner}}
\label{sec:App.RCM.Wegner}

We will need two results from our earlier works.

The first of them is a simple adaptation of a multi-particle EVC bound from \cite{C10}.

\bpr[Cf. \cite{C10}*{Theorem 1}]
\label{prop:Np.EVC.RCM}
Consider a weakly $Q$-separated pair of balls $\bball_L(\Bx)=\bballN_L(\Bx)$,
$\bball_L(\By)=\bballN_L(\By)\subset\bcZN$,
with some $Q \subset\DZ^d$ (cf. \eqref{eq:def.weak.sep}),
and random $\fN$-particle Hamitonians $\BH_{\bball_L(\Bx)}(\tomega)$, $\BH_{\bball_L(\By)}(\tomega)$,
relative to a probability space $(\tOmega, \tfF, \prtild)$
Introduce the sample mean $\xi_Q = |Q|^{-1}\sum_{u\in Q} V(u;\tomega)$, the sigma-algebra
$\tfF_\eta$ generated by the fluctuations $\eta_u = V(u;\tomega) - \xi_Q$,
and denote
$$
\nu_Q(r;\tomega) = \sup_{r\in\DR} \; \essup \prtil{ \xi_Q \in [r, r + t] \cond \tfF_\eta}.
$$
Suppose that
\be\label{eq:cond.RCM.Thm}
\pr{ \nu_Q(t; \cdot) > h_1(t) } \le h_2(t) .
\ee
Then for any $s>0$ the following bound holds:
\be\label{prop.nu.EVC}
\bal
\pr{ \dist\left( \sigma\left(\BH_{\bball_L(\Bx)} \right),
    \sigma\left(\BH_{\bball_L(\By)} \right) \right) \le t }
&\le  h(t)
\eal
\ee
with
\be
h(t) := |\bball_L(\Bx)| \cdot |\bball_L(\By)| \, h_1(t) + h_2(t).
\ee
\epr

For the reader's convenience, we prove this adaptation of Theorem 1 from \cite{C10} in
Appendix \ref{sec:App.prop.Np.EVC.RCM}.

It is to be emphasized that the nature of the probability space in Proposition
\ref{prop:Np.EVC.RCM} can be quite arbitrary, provided that the bound
\eqref{eq:cond.RCM.Thm} holds true.

The following simple bound of the form \eqref{eq:cond.RCM.Thm} was obtained
in our work \cite{C13a}.

\bpr[Cf. \cite{C13a}*{Theorem 2}]
\label{prop:RCM.Unif}
Let $\{V(x;\tomega), x\in Q\}$, $|Q|<\infty$, be independent random variables with
probability distributions $\Unif(J_x)$, where the intervals $J_x$ have length $\ell>0$.
Then the quantity $\nu_Q(t;\tomega)$ introduced in \eqref{prop.nu.EVC}
satisfies
\be\label{eq:prop.nu.unif.2.3}
\pr{ \nu_Q(t; \cdot) > (\ell \eps)^{-1} t } \le |Q|^2 \eps^{2} .
\ee
\epr

\proof[Proof of Theorem \ref{thm:weak.sep.Wegner}]

Consider a pair of balls $\bball_L(\Bx)$, $\bball_L(\By)$, with $|\Bx - \By|\ge 3 \fN L$
and $|\Bx|, |\By| \le L^4$.
By Lemma \ref{lem:3NL.WS}, this pair is weakly separated. The external potential $\BV(\cdot;\om;\th)$,
restricted on either of these balls, has the form
$$
\Bu = (u_1, \ldots, u_\fN) \mapsto g\sum_{j=1}^\fN V(u_j;\om;\th)
= g\sum_{j=1}^\fN v(T^{u_j}\om;\th),
$$
where $u_j\in\ball_{L^4}(0)$, for all $j=1, \ldots, \fN$.

Next, consider the conditional probability measure, induced by $\DP\times\prthp$, given the
sigma-algebra $\fB_L$ figuring in the hypothesis \LVB; recall that conditional on $\fB_L$,
the sub-sample $\{gV(T^u \om;\th), \, u\in\ball_{L^4}(0)\}$ becomes independent, and the random variables
$V(T^u \om;\th)$ have individual uniform distributions $\Unif\big(J_{x,L}\big)$,
where $J_{x,L}=J_{x,L}(\om;\th)\subset\DR$ are intervals of length
(cf. \eqref{eq:lemma.ell.N.lower.bound})
\be\label{eq:ell.L.again}
g a_{\tNL} \ge gL^{- B \ln L},
\;\; B = 400 b A^2/\ln 2 .
\ee
Note that $ga_{\tN(L_j)}$ corresponds to the parameter $\ell$ in Proposition \ref{prop:RCM.Unif},
so
we set for notational brevity
$$
\ell_j := ga_{\tN(L_j)} = g \cdot  L_j^{-b \tN(L_j)}, \;\
\tN(L_j) \in [3 A \ln L_j, 5 A \ln L_j].
$$

Further, fix $j\ge 0$ and denote for brevity $Q = Q_j = \ball_{L_j^4(0)}$.
By Proposition \ref{prop:RCM.Unif} combined with \eqref{eq:ell.L.again}, with
$t = gs$, $\eps\in(0,\ell_j]$
\be
\bal
\pr{ \nu_Q(t) > \ell_j^{-1} \eps^{-1} gs  } &
 = \pr{ \nu_Q(t) > (ga_{\tNL})^{-1} \eps^{-1} gs  }
\\
& = \pr{ \nu_Q(t) > a_{\tNL}^{-1} \eps^{-1} s  }
\\
& \le |Q|^2\, \eps^2
\eal
\ee
Now apply Proposition \ref{prop:Np.EVC.RCM}:

$$
\pr{ \dist\left( \sigma\left(\BH_{\bball_L(\Bx)} \right),
    \sigma\left(\BH_{\bball_L(\By)} \right) \right) \le g s } \le h( gs),
$$
with
$$
h( gs ) \le 3^{\fN d} L^{2 \fN d} a_{\tN(L_j)}^{-1} \eps^{-1} s
+ |Q|^2\, \eps^2 .
$$
Setting $\eps = s^{1/3}$, we obtain
\be
\bal
h( gs ) &\le 3^{\fN d} L_j^{2 \fN d} a_{\tN(L_j)}^{-1} s^{2/3} + |Q|^2\, s^{2/3}
\\
&\le 3^{\fN d} L_j^{2 \fN d} \eu^{ B \ln L_j } s^{2/3} + 3^{2d} L_j^{8d} \, s^{2/3}
\\
& \le C_5(\fN, d) L_j^{(2\fN+4)d + B \ln L_j} s^{2/3}
\eal
\ee
with some $C_5 = C_5(\fN, d)$. At the last step, we used the inequalities
$\fN\ge 2$ and
$a_{\tN(L_j)}\le \eu^{B\ln L_j}$ (cf. \eqref{eq:lemma.ell.N.lower.bound}). This completes the proof.
\qedhere

\section{ Proof of Proposition \ref{prop:Np.EVC.RCM}}
\label{sec:App.prop.Np.EVC.RCM}

\proof
Let $\bball^{(\fN)}_L(\Bx)$, $\bball^{(\fN)}_L(\By)$  be a weakly separated pair of balls
satisfying the conditions \eqref{eq:def.weak.sep}
for some $Q\subset\DZ^d$, $\cJ_1, \cJ_2 \subset \{1,\ldots,\fN\}$ with  $|\cJ_1|=\fn_1 > \fn_2 =|\cJ_2|$.
For brevity, we will omit the superscript $\fN$.
As in Section \ref{ssec:RCM}, introduce the sample mean $\xi=\xi_{Q}$ of $V$ over $Q$ and the fluctuations
$\{\eta_x, \, x\in Q \}$, so that
$$
\forall\, x\in Q\quad V(x;\om;\th) = \xi_Q(\om;\th) + \eta_x(\om;\th),
$$
and let $\fB^Q_\eta$
be the sigma-algebra  generated by
$\{\eta_x^\Lam(\cdot), \, x\in Q; \;V(y;\cdot;\cdot), \, y\in \cZ \setminus Q   \}$
(cf. \eqref{eq:eta.Lam}--\eqref{eq:fB.eta.Lam}).
Then the operator
$$
\BH_{\bball_{L}(\Bx)}(\omega;\th) = \BH_0 + \BU + g\sum_{j=1}^\fN V(x_j;\om;\th)
$$
admits the following representation:
$$
\bal
\BH_{\bball_{L}(\Bx)}(\omega;\th)
&= g\sum_{j\in\cJ_1} V(x_j;\om;\th) + \left(\BH_0 + \BU + g\sum_{j\not \in \cJ_1} V(x_j;\om;\th) \right)
\\
& = g\, \fn_1 \, \xi_Q(\om;\th) \, \one + \BA'(\omega;\th),
\eal
$$
where the operator $\BA'(\omega)$ is $\fF_{Q}$-measurable.
Similarly,
\begin{equation}\label{eq:Ham.decomp}
\begin{array}{l}
\BH_{\bball_{L}(\By)}(\omega) = g\, \fn_2\, \xi(\omega) \,\one + \BA''(\omega),
\; 0 \le \fn_2 < \fn_1,
\end{array}
\end{equation}
where $\BA''(\omega)$ is $\fB^{Q}_\eta$-measurable.
Let
$\{ \lambda_1, \ldots, \lambda_{M'}\}$,
$M' = \,\,|\bball_{L}(\Bx)|$,
and
$\{ \mu_1, \ldots, \mu_{M''}\}$,
$M'' = |\bball_{L}(\By)|$,
be the sets of eigenvalues of $\BH_{\bball_{L}(\Bx)}$ and of $\BH_{\bball_{L}(\By)}$,
counting multiplicity.
Owing to \eqref{eq:Ham.decomp}, these eigenvalues can be represented as follows:
$$
\begin{array}{l}
\lambda_j(\omega) = \fn_1\xi(\omega) + \lambda_j^{(0)}(\omega),
\quad
\mu_j(\omega) = \fn_2\xi(\omega) + \mu_j^{(0)}(\omega),
\end{array}
$$
where the random variables
$\lambda_j^{(0)}(\omega)$ and $\mu_j^{(0)}(\omega)$ are $\fF_{Q}$-measurable. Therefore,
$$
\lambda_i(\omega) - \mu_j(\omega) =  (n_1-n_2)\xi(\omega) + (\lambda_j^{(0)}(\omega) -  \mu_j^{(0)}(\omega)),
$$
with $\fn_1-\fn_2 \ge 1$, owing to our assumption.
Further, we can write
$$
\bal
\pr{ \dist(\Sigma_\Bx, \Sigma_\By)) \le t }
&= \pr{ \exists\, i,j:\, |\lambda_i - \mu_j| \le t }
\\
& \le \sum_{i=1}^{M'} \sum_{j=1}^{M''}
     \esm{ \pr{ |\lambda_i - \mu_j| \le t \,| \fF_{Q}}}.
\eal
$$
Note that for all $i$ and $j$ we have
$$
\bal
\pr{ |\lambda_i - \mu_j| \le s \,|\, \fF_{Q}}
& = \pr{ |(\fn_1 - \fn_2)\xi + \lambda_i^{(0)} - \mu_j^{(0)}| \le s \,| \fF_{Q}}
\\
&\le \displaystyle \nu_L( 2|\fn_1 - \fn_2|^{-1} t \,|\, \fF_{Q}) \le \nu_L( 2 t \,| \fF_{Q}).
\eal
$$
Consider the event
$$
\cE_L = \Bigl\{\; \sup_{r\in\DR} \;
\big|F_\xi(r+t\,| \fF_{Q}) - F_\xi(t\,| \fF_{Q})\big|\ge h_1(t) \Bigr\}.
$$
By hypothesis, $\pr{\cE_L} \le h_2(t)$.
Therefore,
$$
\bal
\pr{ \dist(\Sigma_\Bx, \Sigma_\By)) \le t }
&= \esm{ \pr{  \dist(\Sigma_\Bx, \Sigma_\By) \le t  \,| \fF_{Q}}}
\\
&\le \esm{ \one_{\cE^c_L} \pr{ \dist(\Sigma_\Bx, \Sigma_\By) \le t \,| \fF_{Q}}}
+ \pr{\cE_L}
\\
&\le |\bball_{L}(\Bx)| \cdot |\bball_{L}(\By)|\, h_1(t) + h_2(t) = h_L(t),
\eal
$$
as asserted.
\qedhere

\section{Proof of Lemma \ref{lem:NT.NR.is.NS}}
\label{app:lem.NT.NR.is.NS}

The assertion of Lemma \ref{lem:NT.NR.is.NS} is actually a particular case of a more general statement
concerning the decay properties of $(\ell,q)$-dominated functions on graphs,
which we formulate and prove below.
In turn, this is merely a variant of an older result, appeared first in the works
by von Dreifus \cite{Dr87}, Spencer \cite{Sp88}, and in a modified and generalized form,
in the paper by von Dreifus and Klein \cite{DK89}.

We denote by $\fG(D,C_D)$ the class of connected graphs $\cG$ of with polynomial
growth of balls: $\cG\in\fG(D,C_D)$ iff
$\forall\,x\in\cG\; \card \ball_L(x) \le C_D (1\vee)L^D$, $D<\infty$.

\ble\label{lem:NT.NR.is.NS.gen}
Let $\fG(D,C_D)$ be a connected graph, and
$\Lam\subset \cG$ a finite subgraph such that $\Lam \supset \ball_{2L}(x)$,
$L = \lfloor \ell^\alpha\rfloor$ with $\alpha\ge 3/2$, $\ell \ge 4^8(\cK+1)^4$,
$\cK\in\DN$.
Consider a DSO $H_\Lam = \Delta_{\Lam} + W$ in $\Lam$ with (fixed) potential $W:\Lam\to\DR$.
Fix any $E\in\DR$ and
suppose that there is at most one ball $\ball_{\cK \ell}(w)\subset \ball_{2L}(x)$ such that
every ball $\ball_{\ell}(v)\subset \left(\ball_{2L}(x) \setminus \ball_{\cK \ell}(w)\right)$
is $(E,m)$-NS. If $\dist\big( \Sigma(H_{\ball_L(x)}),E\big) \ge \eu^{-L^\beta}$ with $\beta \le 1/2$,
then $\ball_L(x)$ is $(E,m)$-NS.
\ele

\proof
It suffices to consider the case where the ball $\ball_{\cK \ell}(w)$ figuring in the
hypotheses of the lemma is actually present; otherwise, the argument given below becomes
simpler. We will assume that $\alpha $ is an integer, so $L = \ell^\alpha$;
this is the case in the proof of Lemma \ref{lem:NT.NR.is.NS}. Otherwise, one should
take into account the rounding errors.
\par
\noindent
\textbf{Step 1.} We start by assessing the Green functions $G_{\ball_{2L}(x)}(u,x;E)$ with
$\rd(u,x)\in\{L, L+1\}$.

Assume first that $\rd(u,w) \ge 3\ell$ and $\rd(x,w) \ge 3\ell$. By the triangle inequality,
there exist integers $r', r''$ such that
$
\ell \le r', r'' \le L - 2\ell, \;\; r' + r'' = L - 4\ell,
$
and $\ball_{r'}(x)$, $\ball_{r''}(u)$, $\ball_{\cK \ell}(w)$ are pairwise disjoint. Since
$\rd(x,u)\le L+1$ and $r', r''\le L - 2\ell$, we have
$\ball_{r'}(x), \ball_{r''}(u) \subset \ball_{2L-\ell}(x)$, thus any ball $\ball_\ell(v)$
inside $\ball_{r'}(x)$ and inside $\ball_{r''}(u)$ must be $(E,m)$-NS.
Let $f:(s,t) \mapsto G_{\ball_{2L}(x)}(s,t;E)$, for $s,t\in\ball_{2L}(x)$.
This function is $(\ell,q)$-dominated in $s\in\ball_{r'}(x)$
and in $t\in\ball_{r''}(u)$, with $q=\eu^{-m\ell}$.
Applying Lemma \ref{lem:subh.1},
we can write, therefore, with the convention $-\ln 0 = +\infty$:
$$
\bal
-\ln f(x,u) &\ge -\ln \left\{
\big(\eu^{-m\ell( 1 + \ell^{-1/8} )} \big)^{ \frac{L - 2\cK\ell - 2\ell}{\ell + 1}} \; \eu^{L^\beta}
\right\}
\\
& \ge m( 1 + \ell^{-1/8} ) \ell \left( \frac{L - (2\cK+2)\ell }{\ell + 1} \right) - L^{\beta}
\\
& \ge m( 1 + \ell^{-1/8} ) L \Big( \big(1 -(2\cK+2)\ell^{-\alpha+1}\big)\big(1 - \ell^{-1}\big) - \ell^{-\alpha(1-\beta)}\Big)
\eal
$$
hence with $\alpha \ge 3/2$, $\beta \le 1/2$, and $\ell \ge 4^{8}(\cK+1)^4 \ge 2^{16}$,
we have
\be\label{eq:NT.NR.implies.NS.step1}
\bal
-\ln f(x,u) &\ge
m L ( 1 + \ell^{-1/8} )\Big( \big(1 -(4\cK+4)\ell^{-1/4}\big) - \ell^{-5/8} \Big)
\\
& \ge m L ( 1 + \ell^{-1/8} )\Big( \big(1 -(8\cK+8)\ell^{-1/4}\big) \Big)
\\
& \ge m L \Big( 1 + \half \ell^{-1/8} \Big) \ge m L \Big( 1 + L^{-1/8} \Big).
\eal
\ee
Now let $\rd(u,w)<3\ell$ (the case $\rd(x,w)<3\ell$ is similar), and set $r'=L - 3\ell>L-4\ell$,
$r''=0$. Applying Lemma \ref{lem:subh.1} to the ball $\ball_{r'}(x)$, without using the other ball
$\ball_{r''}(u)$, we obtain the same (indeed, a better) upper bound on $|G_{\ball_L(x)}(x,u;E)|$.

\par
\noindent
\textbf{Step 2.}
Denote
$G^\bullet(E) = \left( H_{\ball_L(x)} \oplus H_{\ball_{2L}(x) \setminus \ball_L(x)} -E\right)^{-1}$.
Then for any $y$ with $\rd(y,x)=L$, we have, by the GRE \eqref{eq:GRE.GF},
$$
\bal
&|G_{\ball_{L}(x)}(y,x;E)| \le \big|G_{\ball_{2L}(x)}(y,x;E) \big|
\\
& \qquad {+ |\pt \ball_L(x)| \, \max_{y,z\in\ball_L(x)} |G^\bullet_{\ball_{L}(x)}(y,z;E)|
\, \max_{\rd(z'x)=L+1} | G_{\ball_{2L}(x)}(z',x;E) | }.
\eal
$$
For all $y,z\in\ball_L(x)$, we have
$G^\bullet_{\ball_{L}(x)}(y,z;E) = G_{\ball_{L}(x)}(y,z;E)$,
Since $\ball_{L}(x)$ is $E$-NR, we have
$|G_{\ball_{L}(x)}(y,z;E)|\le \|G_{\ball_{L}(x)}(E)\|\le\eu^{-L^\beta}$.
Recalling \eqref{eq:NT.NR.implies.NS.step1}, we obtain:
$$
\bal
-\ln|G_{\ball_{L}(x)}(y,z;E)|
& \ge mL\big(1 + {\textstyle \half } \ell^{-1/8}\big) - \ln\left(1 + |C L^D \eu^{L^\beta}| \right)
\\
& \ge  mL\big(1 + {\textstyle \quart} \ell^{-1/8}\big) \ge mL\big(1 + L^{-1/8}\big)
= \gamma(m,L)L,
\eal
$$
provided that $\ell$ (hence, $L/\ell = \ell^{\alpha-1}$) is large enough.
Thus $\ball_L(x)$ is $(E,m)$-NS.
\qedhere

\section*{Acknowledgements}

It is a pleasure to thank
Yakov Grigor'yevich Sinai,
Abel Klein and Misha Goldstein for numerous fruitful discussions of localization mechanisms
in deterministic disordered media;
Tom Spencer for numerous stimulating discussions of the multi-particle localization and of almost-periodic
operators, as well as for his warm hospitality during my stay at the IAS in 2012;
G\"{u}nter Stolz, Yulia Karpeshina and Roman Shterenberg
for stimulating discussions and  warm hospitality during my stay at the University
of Alabama at Birmingham in 2012;
Werner Kirsch and the Fern\-Universit\"{a}t Hagen for their warm hospitality in 2013;
Fumihiko Nakano and the Gakushuin University of Tokyo for the stimulating discussions
and the warm hospitality, as well as
Shinichi Kotani and Nariyuki Minami for fruitful discussions and the
Research Institute for Mathematical Sciences (RIMS, Kyoto) for the warm hospitality
during my visit to Japan in 2013.


\end{document}